\documentclass[structabstract]{aa}
\usepackage{graphicx}
\usepackage{natbib}
\bibpunct{(}{)}{;}{a}{}{,}
\usepackage{txfonts}
\usepackage{color}

\begin{document}
    \title{Pulsation spectrum of  $\delta$ Sct stars:\\
the binary HD 50870 as seen with CoRoT and HARPS
\thanks{
The CoRoT space mission was developed and  is operated by the French
space agency CNES, with participation of ESA's RSSD and Science Programmes,
Austria, Belgium, Brazil, Germany, and Spain.
This work is based on ground--based observations made with ESO telescopes
at La Silla Observatory under the ESO Large Programmes LP~182.D-0356
and LP~185.D-0056 and on data collected at 
the Observatorio Astron\'omico Nacional San Pedro M\'artir (Mexico).}
} 
   \author{  
L.~Mantegazza\inst{1}
\and
E.~Poretti\inst{1} 
\and
E.~Michel\inst{2}
\and
M.~Rainer\inst{1}
\and
F.~Baudin\inst{3}
\and
A.~Garc\'ia Hern\'andez\inst{4}
\and
T.~Semaan\inst{5}
\and
M.~Alvarez\inst{6}
\and
P.J.~Amado\inst{4}
\and
R.~Garrido\inst{4}
\and
P.~Mathias\inst{7}
\and
A.~Moya\inst{8}
\and
J.C.~Su\'arez\inst{4}
\and
M.~Auvergne\inst{2}
\and
A.~Baglin\inst{2}
\and
C.~Catala\inst{2}
\and
R.~Samadi\inst{2}}
\authorrunning{Mantegazza et al.}
\titlerunning{HD 50870 as observed with CoRoT and HARPS}
\offprints{L.~Mantegazza\\
\email{luciano.mantegazza@brera.inaf.it} }

   \institute{INAF -- Osservatorio Astronomico di Brera, Via E. Bianchi 46, 23807 Merate, Italy
              \and
LESIA, UMR 8109, Observatoire de Paris, 5 pl. J. Janssen, 92195 Meudon Cedex, France
\and
Institut d'Astrophysique Spatiale, CNRS, Universit\'e Paris XI UMR 8617, 91405 Orsay, France
\and
Instituto de Astrof\'{\i}sica de Andaluc\'{\i}a (CSIC),
 Glorieta de la Astronom\'{\i}a S/N, 18008 Granada, Spain
\and
GEPI, Observatoire de Paris, CNRS, Universit\'e Paris Diderot; 5 Place Jules Janssen, 92195 Meudon, France
\and
Observatorio Astron\'omico Nacional, Instituto de Astronomia, UNAM, Apto Postal 877,
Ensenada, BC 22860, M\'exico
\and
Institut de Recherche en Astrophysique et Plan\'etologie, UMR 5277, 57 avenue d'Azereix, 65000 Tarbes, France
\and 
Departamento de Astrof\'{\i}sica, Centro de Astrobiolog\'{\i}a (INTA-CSIC), PO Box 78, 28691 Villanueva de la Ca\~nada, Madrid, Spain 
             }
   \date{Received / accepted}
   \abstract 
   {}
{
We present the results obtained with  the CoRoT satellite for HD~50870, a $\delta$ Sct star
which was observed for 114.4~d.
The aim of these observations was to  evaluate the results obtained for HD 50844,
the first  $\delta$ Sct star monitored  with CoRoT, on a longer time baseline.
}
 {The 307,570 CoRoT datapoints were analysed with different techniques.
The photometric observations were complemented over
 15 nights of high-resolution spectroscopy with HARPS on a baseline 
 of  25~d. These spectra were analysed to study the line profile variations and to derive the stellar
 physical parameters. Some $uvby$ photometric observations were also obtained to better
 characterize the pulsation modes.}
{HD~50870 proved to be a low-amplitude, long-period spectroscopic binary system 
seen almost pole-on ($i\simeq21^o$). The brighter component, which also has 
the higher rotational velocity ($v \sin i=37.5$~km s$^{-1}$), is a $\delta$ Sct-type 
variable with a full light amplitude variation of about 0.04~mag. There is a
dominant axisymmetric mode (17.16~d$^{-1}$). Moreover,
there are two groups of frequencies (about 19) in the intervals 6-9 and 13-18~d$^{-1}$,
 with amplitudes ranging from a few mmag to 0.3 mmag.  After the detection of about 250
terms (corresponding to an amplitude of  about 0.045 mmag) a flat plateau 
appears in the power spectrum in the low-frequency region up to about 35~d$^{-1}$.
We were able to detect this plateau only thanks to the short cadence sampling of the CoRoT measurements (32~s).
The density distribution vs. frequency of the detected frequencies seems rule out the
possibility that this plateau is the result of a process with a continuum power spectrum.
The spacings of the strongest modes suggest  a quasi-periodic pattern. 
We failed to find a satisfactory seismic model that simultaneously matches the frequency range, the position 
in the HR diagram, and the quasi-periodic pattern interpreted as a large separation.
Nineteen modes were detected spectroscopically from the line profile variations and associated 
to the photometric ones.
Tentative $\ell,m$ values have been attributed to the modes detected 
spectroscopically. Prograde as well as  retrograde modes are present with
$\ell$ values up to 9. 
There are no traces of variability induced by solar-like oscillations.
}
 {}

    \maketitle
%

\section{Introduction}
$\delta$ Sct stars are multimode low--amplitude, short--period, opacity--driven pulsators in the lower 
part of the Cepheid instability strip.
Recent photometric observations performed by the CoRoT (COnvection, ROtation and planetary Transits)
satellite \citep{baglin} have shown 
that about 1000 significant peaks have to be removed to obtain amplitude spectra that
show noise only \citep{por,garcia}.  In particular, the lowest modes 
tend to constitute a flat plateau with a high--frequency cut--off (e.g., around 30 d$^{-1}$ for HD 50844)
after the simultaneous fit and prewhitening of the strongest ones.
\begin{figure*}[]
\begin{center}
\includegraphics[width=2.0\columnwidth,height=\columnwidth]{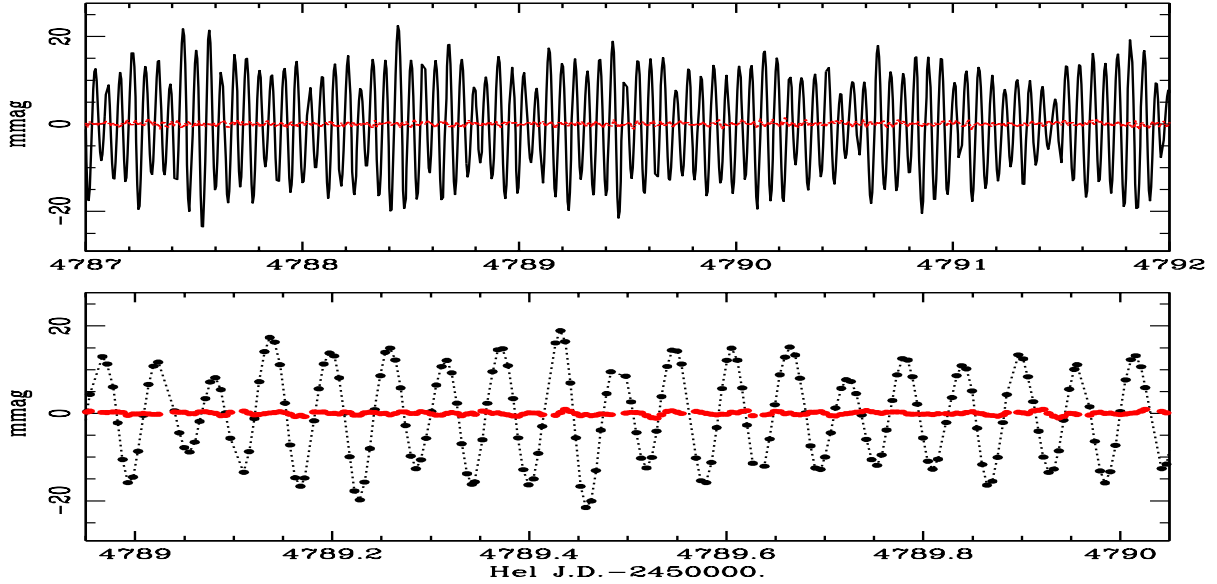} 

\caption{\footnotesize
Examples of the light curve of HD~50870 observed by CoRoT. The lower panel
is a zoom--in of a region of the upper panel. The red points clustering around
zero are the residuals after subtracting the 734 detected frequencies. 
}
\label{lc}
\end{center}
\end{figure*}

The physical interpretation of this  fact leads  to two main possibilities. 
This bunch of peaks could be  explained by the excitation of a large number 
of  modes with a high $\ell$ combined with  an insufficient cancellation effect of the flux 
variations when integrated over the whole stellar disk \citep[HD~50844, ][]{por,dasdas}. 
In this respect, \citet{lgn09} showed how rotation enhances the non-cancellation effects,
finding that there are three times more visible modes in a star rotating
at $\omega$~=~0.59\,$\omega_K$ than in a non-rotating star.
Another possibility is that the plateau is generated by the granulation noise induced by 
the surface convection \citep{kama}.
In connection with the possible presence of a convective envelope, \citet{antoci} reported the
detection of solar--like oscillations in the $\delta$ Sct star HD 187547 observed by {\it Kepler}.
\citet{baldzi} analysed the {\it Kepler} data of  $\delta$ Sct stars 
and  found that in general the frequency density is quite moderate. They suggest  that the high  
density of the frequencies in the HD~50844 data may be an exception.

Additional detailed analyses of the light variation of $\delta$ Sct stars are  necessary  to
clarify these open questions. The aim of the present study is add another tile to the
assembling mosaic by analysing the CoRoT measurements of HD~50870 \citep[$V$=8.88, 
$B-V$=+0.27, F0;][]{mccus}, a $\delta$ Sct discovered in the CoRoT  preparatory work \citep{gaudi}.
To achieve the most complete picture of HD~50870 we complemented the space photometry
with  ground-based high-resolution spectroscopy and with $uvby$ photometry.

\section{The CoRoT data and their frequency content}
CoRoT monitored HD 50870 in 
the second long run in the direction of the anticentre (LRa02),
 started on November 13, 2008 and finished on March 8, 2009   
($\Delta$T=114.41~d). 
For the analysis, we used the reduced N2 data rebinned at 32~s
and only considered the 307,570  datapoints for which no problem (i.e., flag=0) was
reported.
The amplitudes of the background orbital variations were kept at
a  very small level by the great effectiveness of the baffle,  
and the subsequent rejection of the
uncertain points strongly minimized the orbital effects. 
The light curve  was detrended with a low--degree polynomial fit to remove the effect of ageing 
\citep{flight}.

The original data binning of 32~s is higher than necessary for the relevant frequency range
of $\delta$ Scuti stars and therefore we grouped the original data
into new bins with a time separation of 0.005~d to gain CPU time, thus obtaining 20535 datapoints 
and a Nyquist frequency of 100~d$^{-1}$.
An example of the resulting light curve is shown in Fig.~\ref{lc}.

\begin{figure}[]
\begin{center}
\includegraphics[width=\columnwidth,height=\columnwidth]{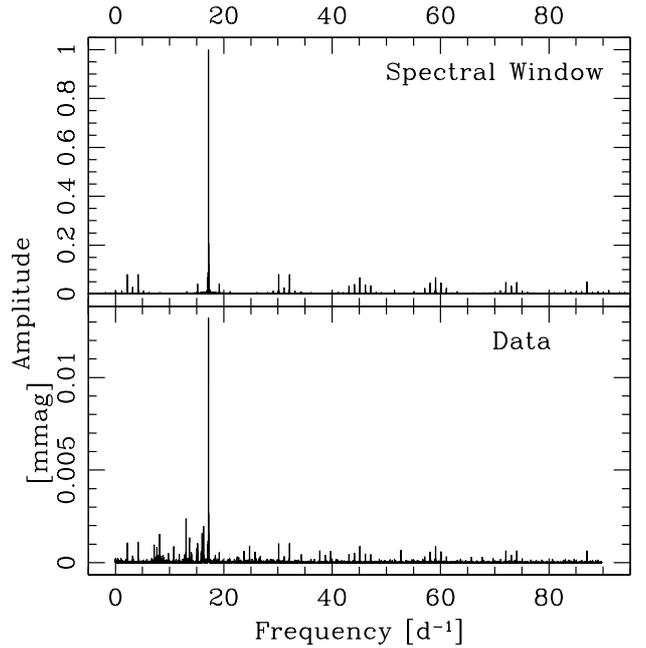}
\caption{\footnotesize {\it Top panel}:
the spectral window of the CoRoT data of HD~50870. 
{\it Bottom panel}: the amplitude  spectrum of the original CoRoT data after the removal of the slow trend.
 }
\label{win}
\end{center}
\end{figure}

The spectral window is shown in Fig.~\ref{win}. The positions of the alias peaks is very similar
to those in the window of HD 50844; their nature has been carefully discussed by 
\citet{por}. Note also  how the amplitude spectrum reflects the spectral window 
above 40~d$^{-1}$, because of the predominant term at 17.16~d$^{-1}$.

The detection of the  frequencies  contributing to the light variations was performed
iteratively by computing the Fourier power spectrum \citep{deeming},
selecting the highest peak, adjusting its
frequency by a non-linear least-squares fit of all previously detected sinusoidal terms plus
the new one, computing the residuals produced by this fit with respect to the original time
series and proceeding to the detection of a new term.
At each step the amplitude signal-to-noise ratio (S/N) was computed for the new detected term 
with respect to the average noise
in a frequency range of 5~d$^{-1}$ centred around its frequency.
The iterations were stopped when a local S/N of 3.0  was  reached.
This approach is similar to that performed by  the Period04 code \citep{P04},
but it is made by means of a completely automated f77 code, which also allows
detecting  more frequencies (Period04 is limited to about 200 components).
The frequency analysis was performed using the relative fluxes, then the results were scaled into mmag units.
\begin{figure}[]
\begin{center}
\includegraphics[width=\columnwidth,height=\columnwidth]{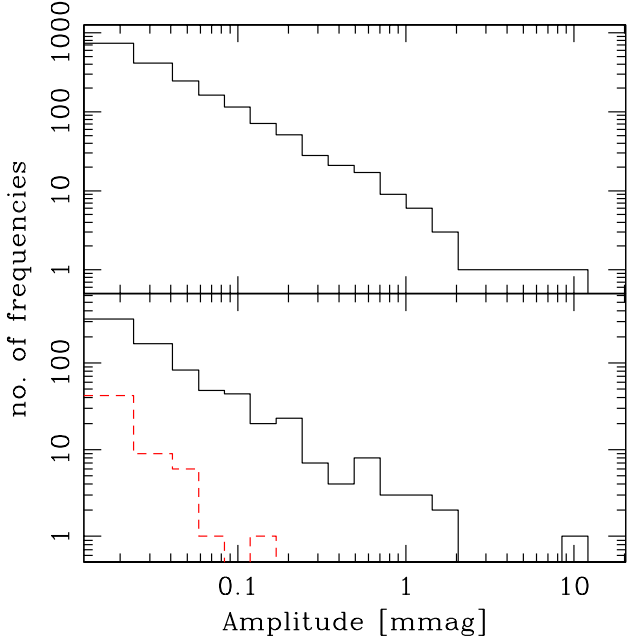}
\caption{\footnotesize 
Distribution of the frequencies vs. amplitudes.
{\it Top panel:} Number of frequencies with an amplitude higher than a given value.
{\it Bottom panel:} histogram of the  734 frequencies (black solid line); distribution of the 
 frequencies that were not detected by SigSpec (red dashed line). Note the logarithmic scales.
}
\label{isto}
\end{center}
\end{figure}

\begin{figure}[]
\begin{center}
\includegraphics[width=\columnwidth,height=\columnwidth]{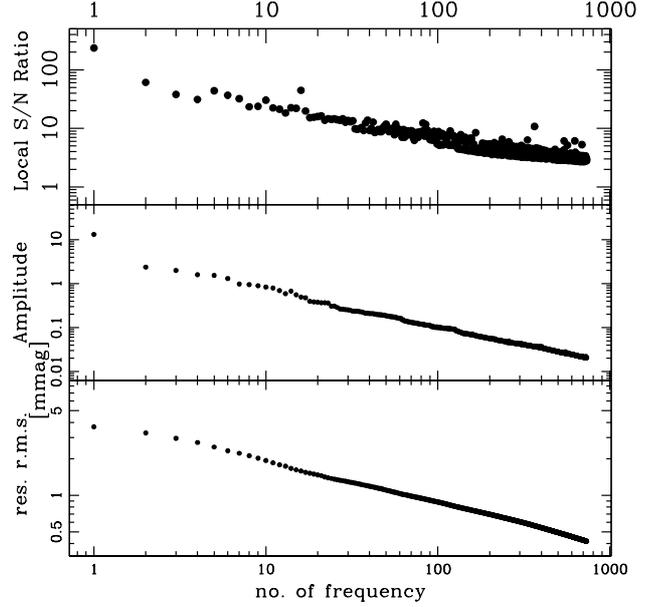}
\caption{\footnotesize
Evolution of the frequency analysis of the CoRoT time series.~{\it Top panel:} local S/N 
of the n$^{th}$ highest peak. {\it Middle panel:} amplitude of the n$^{th}$ highest peak.~ {\it Bottom panel: }
residual r.m.s. 
after subtracting all detected frequencies up to n$^{th}$ term. Note the logarithmic scales on
both axes.
In the top panel the two points with the largest deviations are the Fourier harmonics of the dominant mode.
}
\label{fourisult}
\end{center}
\end{figure}

\begin{table}
\begin{flushleft}
\caption{First 24 frequencies identified in the amplitude spectrum (those with an amplitude $>$0.3 mmag).
The last colum indicates  possible combination terms also present in the data.
}
\begin{tabular}{r rrr c}
\hline \hline
\noalign{\smallskip}
\multicolumn{1}{c}{Term} & \multicolumn{2}{c}{Frequency}&
\multicolumn{1}{c}{Ampl.} & Possible combination  \\
\multicolumn{1}{c}{}& \multicolumn{1}{c}{[d$^{-1}$]} & 
 \multicolumn{1}{c}{[$\mu$Hz]} & 
\multicolumn{1}{c}{[mmag]} &terms \\
\noalign{\smallskip}
\hline
\noalign{\smallskip}
 $f_1$    & 17.1617  &  198.630  &  12.961 & $f_{16}=2f_1$, $f_{365}=3f_1$ \\
 $f_2$    &13.0499  &  151.040  &   2.342 & $f_{87}=f_1+f_2$, $f_{104}=f_1-f_2$\\
          &         &           &         & $f_{726}=2f_2$ \\
 $f_3$    &16.2507  &  188.086  &   1.736 & $f_{82}=f_1+f_3$ \\
 $f_4$    &16.0024  &  185.212  &   1.593 & $f_{85}=f_1+f_4$, $f_{150}=f_1-f_4$ \\
 $f_5$    & 8.1366  &   94.173  &   1.544 & $f_{96}=f_1+f_5$, $f_{447}=2f_5$, \\
          &         &           &         & $f_{597}=3f_5$\\
 $f_6$    &13.6605  &  158.107  &   1.313 & $f_{95}=f_1+f_6$, $f_{190}=f_1-f_6$\\
 $f_7$    & 7.1320  &   82.546  &   0.974 & $f_{356}=f_1+f_7$, $f_{554}=2f_7$, \\
          &         &           &         & $f_{721}=3f_7$ \\
 $f_8$    &15.1787  &  175.678  &   0.979 & $f_{136}=f_1+f_8$\\
 $f_9$    &17.2805  &  200.005  &   0.903 & $f_{167}=f_1+f_9$, $f_{187}=f_9-f_1$\\
 $f_{10}$ & 7.6604  &   88.661  &   0.824 & $f_{115}=2f_{10}$, $f_{178}=f_1+f_{10}$\\
 $f_{11}$ &15.0390  &  174.061  &   0.836 & \\
 $f_{12}$ &16.2473  &  188.046  &   0.819 & \\
 $f_{13}$ &17.1863  &  198.914  &   0.692 & \\
 $f_{14}$ &15.8344  &  183.267  &   0.726 & \\
 $f_{15}$ &15.8217  &  183.120  &   0.707 & \\
 $f_{16}$ &13.9725  &  161.718  &   0.792 & \\
 $f_{17}$ &34.3232  &  397.257  &   0.501 & \\
 $f_{18}$ &14.0608  &  162.740  &   0.471 & \\
 $f_{19}$ & 8.7541  &  101.320  &   0.390 & \\
 $f_{20}$ & 7.8843  &   91.253  &   0.384 & $f_{269}=2f_{20}$, $f_{284}=3f_{20}$\\
 $f_{21}$ &18.4425  &  213.453  &   0.377 & \\
 $f_{22}$ & 8.2945  &   96.001  &   0.371 & \\
 $f_{23}$ & 7.5452  &   87.328  &   0.324 & $f_{60}=2f_{23}$\\
 $f_{24}$ & 8.8886  &  102.877  &   0.308 & $f_{70}=2f_{24}$, $f_{505}=3f_{24}$\\          
\noalign{\smallskip}
\hline
\label{comb}
\end{tabular}
\end{flushleft}
\end{table}

\begin{figure}[]
\begin{center}
\includegraphics[width=\columnwidth,height=\columnwidth]{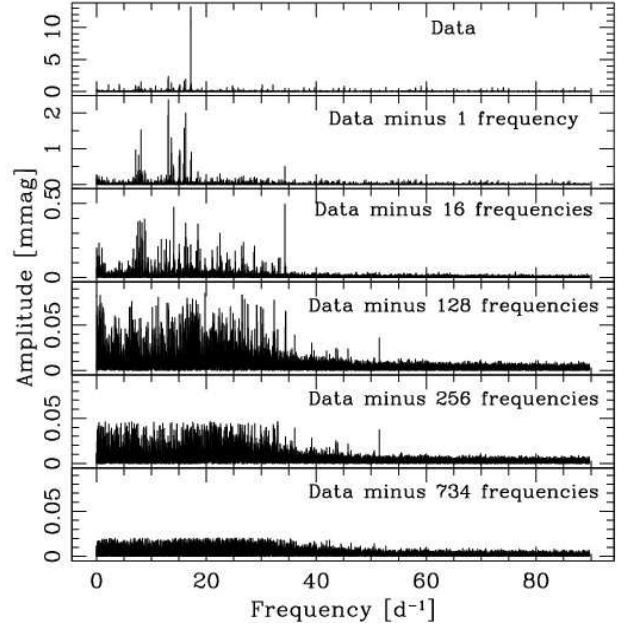}
\caption{\footnotesize
Fourier amplitude spectra at different stages of the analysis.
The flat plateau below about 35~d$^{-1}$ appears after the detection of about 250 frequencies.
The highest peak in the third panel from top is the second harmonic of the dominant frequency,
while its third harmonic is the peak at the highest frequency in 4$^{th}$ and 5$^{th}$ panel.
}
\label{fourier}
\end{center}
\end{figure}

We detected 734 frequencies with amplitudes ranging from 12.961~mmag down to
0.020~mmag. 
The distribution of the detected frequencies vs. their amplitudes is shown in Fig.~\ref{isto}. 
The developments of the analysis are plotted in Fig~\ref{fourisult}, where we can see step by step the
amplitudes of the detected terms, their local S/N at the detection, and the decrease of the residual rms.
These plots  allow us to see how many components are needed to go below a certain level
of S/N or rms residual or amplitude.

Figure~\ref{fourier} shows how the amplitude spectrum changes its shape at the  various stages of the analysis.
At first one strong dominant peak is visible at $f_1$=17.1617~d$^{-1}$, then two groups of peaks show up in the
ranges 7--9~d$^{-1}$ and 12--18~d$^{-1}$. Successively, bunches of peaks appear between 0 and 35~d$^{-1}$.
It is only after the detecting of about 250 terms that the 0--35~d$^{-1}$ region becomes a flat plateau,
progressively decreasing in height after performing new peak detections.
The dominant peak at about 34~d$^{-1}$ in the third panel from the top is the 2$f_1$ harmonic,
while the isolated peak at about 51~d$^{-1}$ in the fourth and fifth panels from the top is 
the 3$f_1$ harmonic.
No higher-order harmonic of $f_1$ was detected.
The white noise level is 0.0031~mmag as estimated in the region  60--90~d$^{-1}$, which leads 
to S/N=6.7 for the plateau below  $\sim$35~d$^{-1}$ visible in the bottom panel of Fig.~\ref{fourier}.

The components with an amplitude higher than 0.3~mmag (24 frequencies) are listed in Table~\ref{comb}. 
Among the detected peaks $f_{16}$ (13.9725~d$^{-1}$) corresponds to the satellite orbital frequency, 
while $f_{418}$ (2.0029~d$^{-1}$, 0.037 mmag) could be produced by the gaps
introduced in the timeseries by the  transit
of the satellite through the South Atlantic Anomaly (SAA), and therefore they are probably not
 attributable to the star.
In the solution of HD~50870 we find some frequencies with a high rank of
detection that exactly match 
the linear combination between a lower-rank frequency and  
the first, predominant term $f_1$ (last column in Table~\ref{comb}).
They probably arise  from nonlinear amplitude distortion
and are commonly found in multiperiodic radial pulsators (Cepheids, RR Lyr, high-amplitude
$\delta$ Sct stars) as well as in low--amplitude $\delta$ Sct stars, e.g. FG Vir \citep{fgvir}. 
The sums and, in some cases, the differences are detected for the modes up to $f_{10}$.
The identification of some high--rank frequencies with the harmonics of low--amplitude
terms ($f_{20}$, $f_{23}$, and $f_{24}$) is reported in Table~\ref{comb} as a mere 
possibility, since they are embedded in the flat plateau below 35~d$^{-1}$ and an accidental
agreement could occur.

\begin{figure}[]
\begin{center}
\includegraphics[width=\columnwidth,height=\columnwidth]{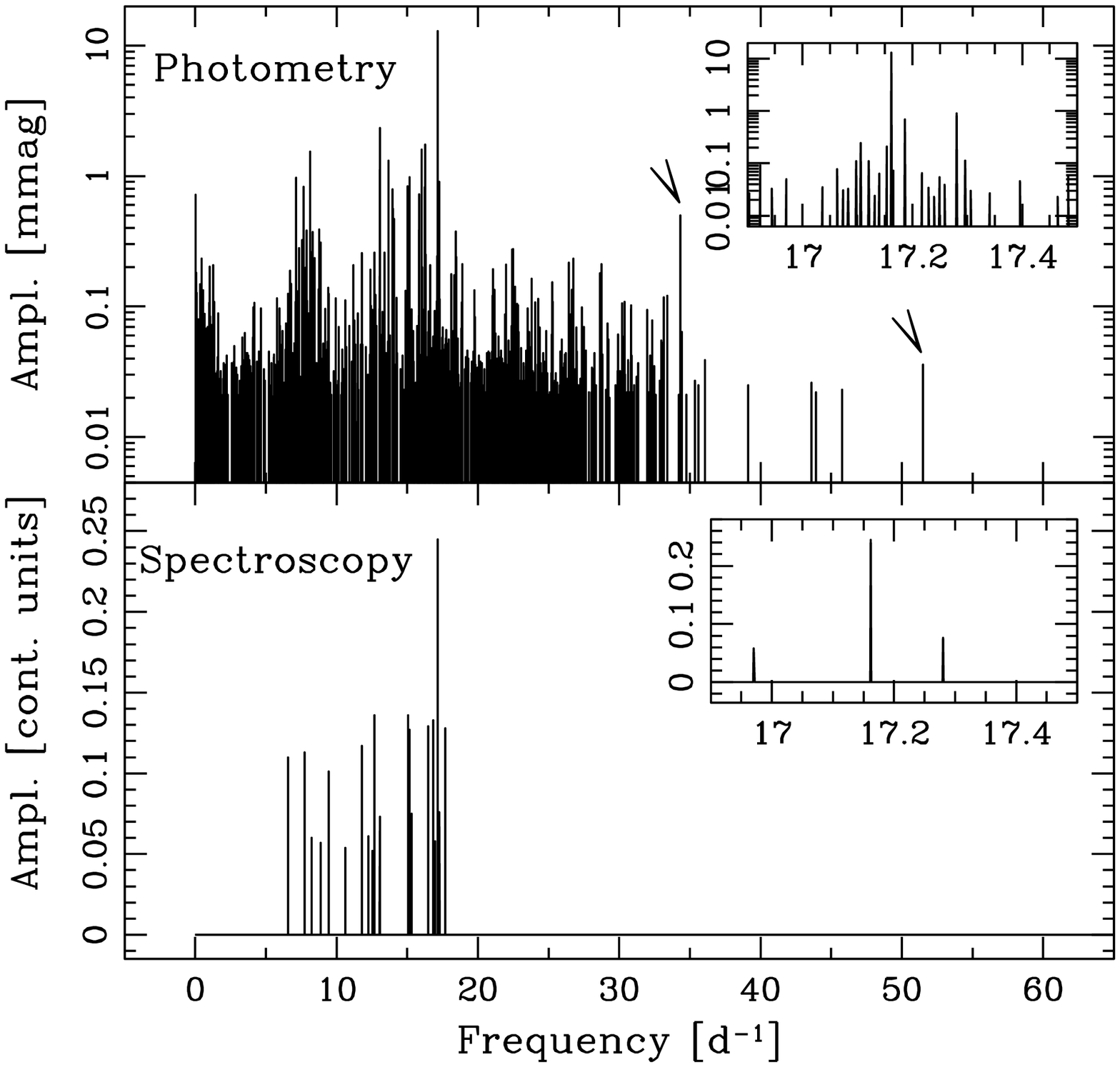}
\caption{\footnotesize
Distribution of the  734 frequencies detected in the CoRoT data (top panel)
and of the 19 detected in spectroscopy  below the pseudo-Nyquist frequency
of 21.06~d$^{-1}$ (bottom panel).
In the top panel the two arrows show the second and third harmonics of the predominant 
mode.  The insets show a zoom around the predominant frequency.
}
\label{freqord}
\end{center}
\end{figure}

The data were also analysed with the SigSpec code \citep{sigspec}.
To go below  S/N=4.0 (corresponding to a value of the $sig$ parameter of about 5.46),
SigSpec requires 790 frequencies.  The  difference in the number of significant peaks 
with respect to the Fourier analysis is probably due to the different
approaches in computing the white noise level.
When comparing the two approaches, we found that 673 of the 734 frequencies detected with the Fourier analysis
are at a distance shorter than 0.0044~d$^{-1}$ (half the formal frequency resolution) from the 
corresponding  frequency detected with SigSpec. In turn, this means that
61 frequencies detected with Fourier were not detected with SigSpec. The first missed detection is the 141-th 
frequency, whose amplitude is 0.090~mmag. The histogram with the detections 
missed by SigSpec vs. the Fourier amplitudes is shown in Fig.~\ref{isto}. 
 Since the 734  detected frequencies are concentrated in the interval 0--35~d$^{-1}$, the probability of 
randomly finding two independent frequencies in the same resolution bin with the two approaches is 
less than 0.2.

To check the reliability of the detected frequency terms with the highest amplitude, 
the dataset was subdivided into two
equal subsets, each with a length of  half  the original baseline.
Each subset was  analysed independently and the detected frequencies were compared 
with those obtained from the whole dataset. 
Of the first 150 frequencies detected from the whole dataset 146 were independently 
detected from the two subsets.  The missing components are the number 100, 129, 133, and 137. 
Component number 100 has a frequency of 13.0601~d$^{-1}$,
and is therefore not resolved in the subsets from the second (13.0499~d$^{-1}$). 
The other three missing frequencies were in any case
independently detected in one of the two subsets.

The very recent result by \citet{antoci} concerning the detection of solar--like oscillations
in the $\delta$ Sct star HD~187547 prompted us to look carefully at the power spectra of HD~50870.
HD~187547 shows  solar-like frequencies in the range of 43-73~d$^{-1}$ (i.e., 500-850 $\mu$Hz) 
with amplitudes up to 0.15 mmag, while the $\delta$ Sct oscillations are confined to
frequencies lower than about 48~d$^{-1}$ (550 $\mu$Hz). 
 The physical parameters of the two stars, as 
estimated from their spectra and multicolour photometry, are quite similar, which
is also supported by the frequency ranges covered by the highest-amplitude modes in both stars.
 We should therefore expect to find solar--like frequencies in the same region.
For HD~50870 all detected frequencies are confined below 51.5~d$^{-1}$ ($\sim$600~$\mu$Hz).
No significant peak is present above this limit. We estimated the white noise level there to be at
about 3~ppm. Only five of the detected frequencies above 35~d$^{-1}$ have amplitudes higher
 than 20~$\mu$mag. One of them is the third harmonic of the predominant mode and the others are still close 
to the limit of 35~d$^{-1}$. 
 Moreover, a frequency analysis performed on the data with the original binning of 32~s
showed that in the frequency region between 90 and 500~d$^{-1}$ there are no peaks with an amplitude
higher than 6~$\mu$mag.  Since there is no regular spacing, the frequency analysis of the CoRoT data
on  HD~50870 does not supply any hint of the excitation of solar--like oscillations at first sight.
\begin{figure}[]
\begin{center}
\includegraphics[width=\columnwidth,height=\columnwidth,angle=90]{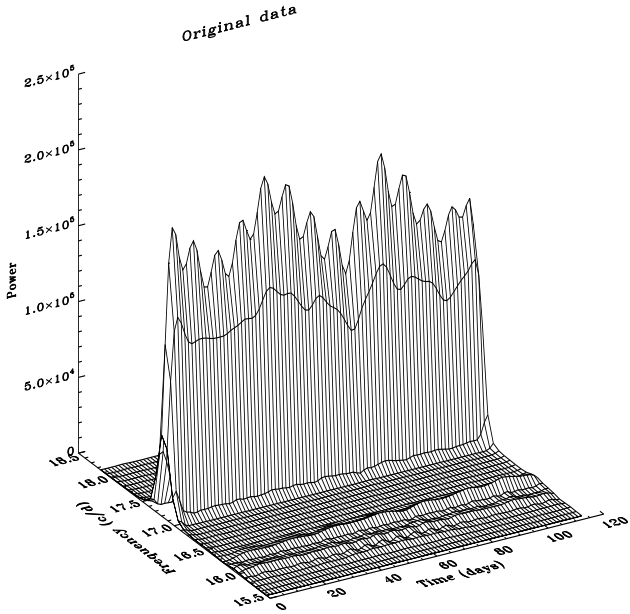}
\caption{\footnotesize
Amplitude behaviour vs. time of the oscillations in the frequency range [15.5,18.5]~d$^{-1}$ 
(i.e., around the dominant mode).
}
\label{baudin1}
\end{center}
\end{figure}

 \begin{figure}[]
\begin{center}
\includegraphics[width=\columnwidth,height=\columnwidth,angle=90]{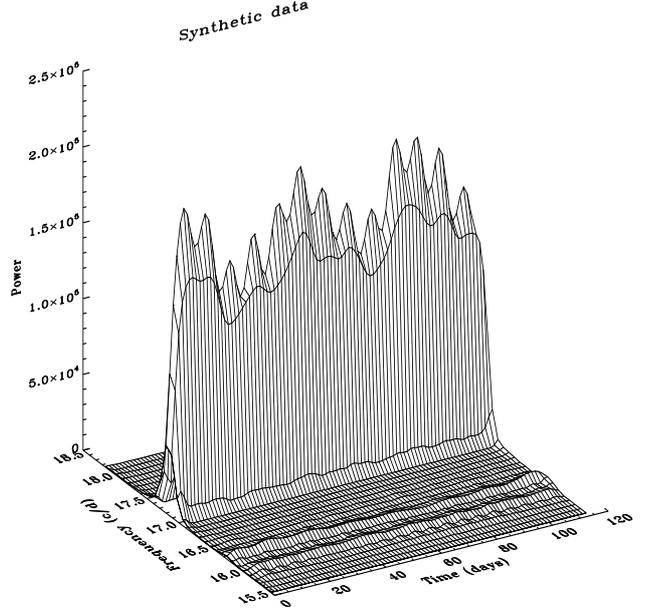}
\caption{\footnotesize
Amplitude behaviour vs. time of a synthetic signal built with the frequency terms detected in the region of the dominant mode
keeping a constant amplitude: there is an excellent agreement with the behaviour derived from the observations.
}
\label{baudin2}
\end{center}
\end{figure}

\subsection{Amplitude stability}

It is rewarding to investigate the stability of the amplitude of the predominant mode 
$f_1=17.17$~d$^{-1}$. 
To do this, we need to consider that the modes $f_9=17.28$~d$^{-1}$ and
$f_{13}$=17.19~d$^{-1}$ are close to $f_1$.
Figure~\ref{baudin1} shows the variation of amplitude recorded when the amplitude modulation
is calculated with  a frequency resolution 
of 0.17~d$^{-1}$, which is insufficient to resolve the three components.

The modulation of the amplitude is clear, but spurious,  because it results from
the beating of the unresolved components. 
We built a synthetic time series using the constant values of amplitudes, frequencies, and phases 
given in Table~\ref{comb} for the three components and sampled it in the same way as the CoRoT data. 
The time-frequency analysis of this synthetic signal shows the same behaviour  of the
amplitude (Fig.~\ref{baudin2}), supporting the conclusion that the amplitudes of the $f_1$,
$f_9$, and $f_{13}$ modes remain constant over the entire time interval covered by the CoRoT observations.

Modes with a lower amplitude were also analysed searching for the typical random variations of 
amplitude and phase of stochastically excited (solar-like) oscillations, but none showed 
this behaviour.

\begin{figure}[]
\begin{center}
\includegraphics[width=\columnwidth,height=\columnwidth]{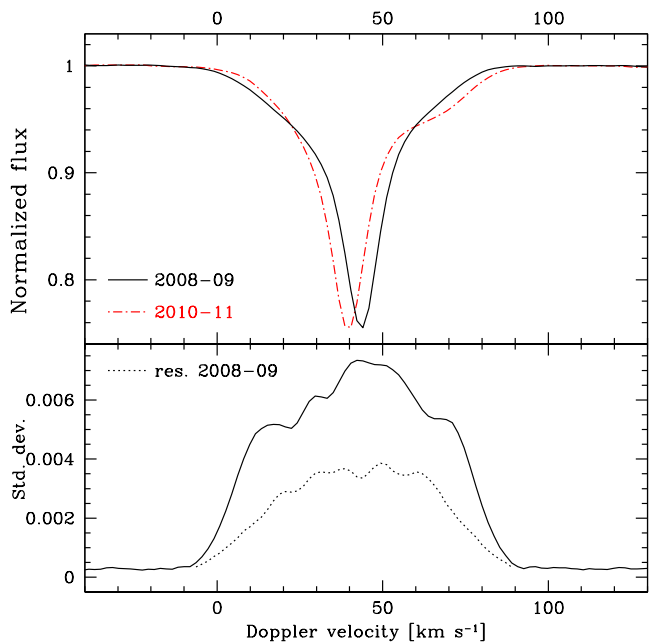}

\caption{\footnotesize
{\it Top panel:} average LSD mean profiles of the 2008-09 (black solid line) and of the 
2010-11 spectra (red dash-dotted line).
{\it Bottom panel:} standard deviations of the original (solid line) 
and of the residual (dotted line) 2008-09 LSD profiles.}
\label{profmed}
\end{center}
\end{figure}

\section{Ground--based observations}
\subsection{High--resolution spectroscopy}
The main batch of spectroscopic observations was performed during 15 nights between December 15, 2008 and
January 8, 2009 with the HARPS instrument mounted on the {3.6-m} ESO telescope, La Silla, Chile.
The spectrograph was used in the high-efficiency EGGS mode, with resolution $R\sim 80,000$.
 We obtained 209 spectra in two runs (ten and five nights, respectively),
 covering a baseline of 24.3~d, i.e., 4.7~times 
shorter than  that of  CoRoT. The exposure time was set to 1200~s and
the S/Ns ranged from 87 to 175 with a median value of 145.  The spectra were reduced
using a semi-automatic pipeline, written in MIDAS and developed by \citet{monica}.
To study the spectral variations induced by pulsation
with the highest S/N possible, we computed the mean line profiles
using the least--squares deconvolution (LSD) method \citep{lsd} 
in the spectral region between 4140 and 5615~$\AA$,
 taking care to omit the intervals  containing the $H_{\beta}$ and  $H_{\gamma}$  lines. 
The resulting LSD profiles have S/Ns ranging from 1260 to 3900 with a mean value of 2406,
i.e., the original S/N was increased by a factor of about 17.
In Fig.~\ref{profmed} we show the average LSD profile (top panel, black solid line) and the standard deviation 
of the original LSD profiles (bottom panel, black solid line).

\subsection{Evidence for binarity}

The average profile shows the superimposition of two components, one narrow and 
another broader. 
The  close scrutiny of the narrow absorption peak 
did not show any appreciable movement with respect to the broader component during the baseline of 25~d.
A non-linear least-squares fit of two Gaussian
components was  performed on the average profile: the centre of the narrow component 
is shifted by about 2.7~km~s$^{-1}$ towards the red edge of the spectrum with 
respect to the centre of the broad one. 

To clarify the possible binary nature of this object nine new HARPS spectrograms were taken between 
December 25, 2010 and January 8, 2011  and three more between December 18, 2011 and
January 12, 2012. The LSD profiles were obtained as described above and their average
shows that  while there is no significant difference between the 2010-11 and 2011-12 spectra,
the two components have moved compared to the positions in 2008-09 (Fig.~\ref{profmed}, top panel).
The narrow component is shifted by 5.2~km~s$^{-1}$ towards the blue edge in the most recent spectra. 
There is also a single FEROS spectrogram ($R$=48000) taken with the FEROS spectrograph attached to the
2.2~m-ESO/MPI  telescope on January 18, 2003: its LSD profile shows
that the narrow component is shifted by 0.9~km~s$^{-1}$ towards the red edge.
We can conclude that we observe a spectroscopic binary.

\begin{figure}[]
\begin{center}
\includegraphics[width=\columnwidth,height=\columnwidth]{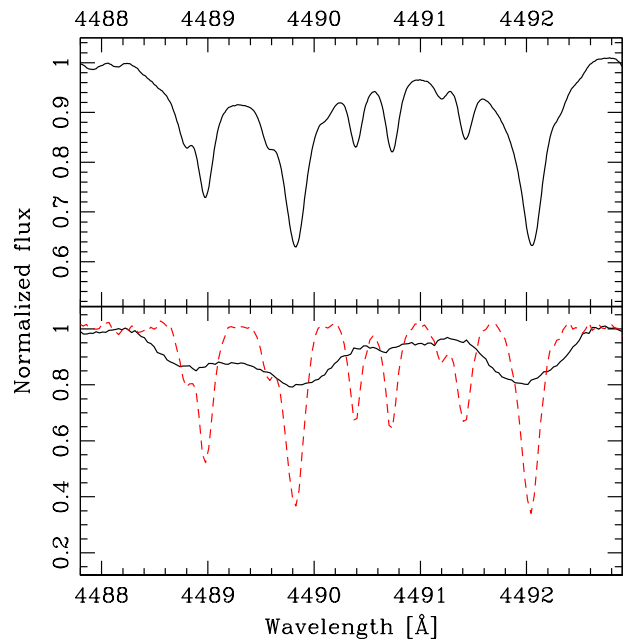}
\caption{\footnotesize
Example of a small  region of the observed spectrum of HD~50870 (top panel) and its separation by the CRES
code in the spectra of the two components (bottom panel).
}
\label{cres}
\end{center}
\end{figure}

We disentangled the spectra with the CRES code \citep{cres} 
to obtain a reasonably accurate estimate of the physical parameters of the two components. 
To do this  we selected some spectral regions that contained lines that are sensitive to the temperature only
 or/and to the gravity only
in  the FEROS spectrogram  and in the two average HARPS spectrograms.
 These spectra were then normalized with the approach described by \citet{ihpf}.
As an example, Fig.~\ref{cres} shows the 2008-09 average spectrum in the wavelength range 4487--4492~$\AA$ and the 
two extracted components.

\begin{figure}[]
\begin{center}
\includegraphics[width=\columnwidth,height=\columnwidth]{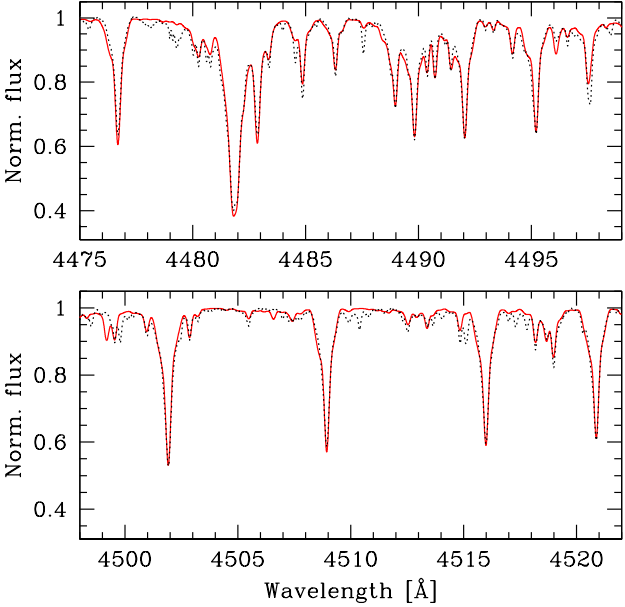}
\caption{\footnotesize
Example of a  comparison between the observed spectrum (black dashed line) and the combination of the
synthetic spectra of the two components obtained with the binmag2 code (red solid line). 
}
\label{fitspec}
\end{center}
\end{figure}

Subsequently,  the physical parameters of the two stars were estimated with the SME code \citep{sme}, 
which fits the observed spectrum with a synthetic one  derived from an interpolation of a grid of models.
Estimates of the effective temperatures, gravities, metallicities, microturbolent
velocities, and projected rotational velocities were obtained by averaging the result of the different
spectral regions (Table~\ref{physpar}).
  Finally, we used the ATC code \citep{atc}  to compute two synthetic spectrograms 
with these parameters and
combined them with the BinMag IDL visualization tool (constructed by O.~Kochukov)
and  compared the combined spectrum with the observed average spectrograms (Fig.~\ref{fitspec}).
The Vienna Atomic Line Database VALD \citep{kupka, vald} was used throughout.
The ratio between the luminosities of the two components at 4500~$\AA$ is  2.0$\pm0.4$.
These results should be considered as preliminary only: additional observations that 
cover more of the orbital period will be able supply firmer results.

\begin{figure}[]
\begin{center}
\includegraphics[width=\columnwidth,height=\columnwidth]{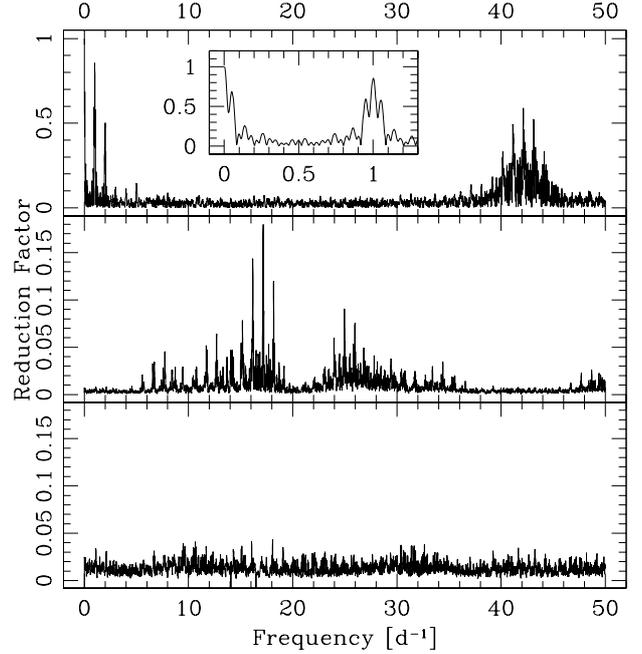}
\caption{\footnotesize
{\it Top panel:} Spectral window of spectroscopic time-series. The high-resolution view of the fine
structure of the peaks is shown in the insert.
{\it Middle panel:} Least-squares power spectrum of LPVs of the original data. The ordinatae show the percentual
data variance reduction.
{\it Lower panel:} Least-squares power spectrum with the 19 detected frequencies given as known constituents.
The ordinates show the percentual reduction of the residual data variance after the fit with the known
frequencies.
}
\label{powersp}
\end{center}
\end{figure}

\begin{table}
\begin{flushleft}
\caption{Physical parameters of the two components}
\begin{tabular}{lcccc}
\hline
\hline 
\noalign{\smallskip}
\multicolumn{1}{c}{} & \multicolumn{1}{c}{Star 1}&
\multicolumn{1}{c}{Star 2} &\multicolumn{1}{c}{Uncertainty} \\
\noalign{\smallskip}
\hline
\noalign{\smallskip}
$T_{\rm eff}$ [K] & 7660 & 7077 & $\pm$250 \\
Spectral type & \multicolumn{1}{c}{A8 III}& \multicolumn{1}{c}{F2 III}\\
$\log g$ & 3.68 & 3.74 & $\pm$0.25 \\
{[Fe/H]} & 0.0 & 0.0 & $\pm$0.2 \\
$v_{mic}$ [km s$^{-1}$]& 3.3 & 2.8 & $\pm$1.0\\
$v\,\sin$~i [km s$^{-1}$]& 37.5 & 8.0 & $\pm$2.5\\
\noalign{\smallskip}
\hline
\label{physpar}
\end{tabular}
\end{flushleft}
\end{table}

\subsection{Line profile variations}

The detection of pulsation modes in the line profile variations (LPVs) was 
performed both with the FAMIAS code  \citep{famias} 
and with the least-squares technique described by \citet{manvie} 
 using the pixel-by-pixel approach. No LPVs were  detected in the 
component with the  narrow lines. On the other hand, clear LPVs were
detected in the component with the broad lines.
Figure~\ref{powersp} shows the least-squares power spectrum obtained from the LSD profiles (top panel)
and the spectral window of the spectroscopic data.
We clearly note that the shorter baseline of the spectroscopic
observations (24~d) implies a frequency resolution worse than that of the
CoRoT photometric data. Moreover,  there are aliases at $\pm$1~d$^{-1}$ from the true peak 
due to the single-site observations, and because the observations were performed in two 
separated runs, each spectral peak is flanked by two symmetrical peaks 
at $\pm$0.051~d$^{-1}$ (Fig.~\ref{powersp}, inserted panel in the top panel).
This hampers a comparison of the photometric frequencies 
with the spectroscopic ones, since  in some cases there are several photometric
frequencies that might be associated to a single spectroscopic peak. This clearly  entails
more uncertainty on the correct identification of the modes.
Moreover, because of the spacing between adjacent spectra, the pseudo-Nyquist frequency
is 21.06~d$^{-1}$. This explains the bunch of peaks around 42~d$^{-1}$ in the spectral
window (Fig.~\ref{powersp}, top panel) and the peaks around 25~d$^{-1}$ and above 48~d$^{-1}$ in 
the first power spectrum (middle panel).
Because of this, attention was paid to avoid misidentifications
of reflected frequency peaks both above and below this value.
The comparison of the findings with the two approaches  
allowed us to identify with some confidence  19 oscillation frequencies (Table~\ref{spectro}).
All frequencies were associated to a photometric term (first column).
There are some ambiguities owing  to the different frequency resolution of the
 photometric and spectroscopic data.  Where that was  the
spectroscopic term was associated to the strongest photometric one that is within the 
HWHM of the centre of the spectroscopic peak. 

The second and third panels of Fig.~\ref{powersp} show the pixel-by-pixel least--squares 
power spectra of LPVs before and after the frequency detection of the 19 above quoted terms.  
Figure~\ref{profmed} (bottom panel, dashed red line) shows the standard deviations across the LSD line profile of the 
residuals obtained by subtracting all the detected components. Evidently,
many undetected components are still present. 

\begin{table}
\caption{Mode detected by the line profile variations listed in order of detection.
All of them were associated to a photometric term.}
\begin{tabular}{r rrrr rr}
\hline 
\hline
\noalign{\smallskip}
\multicolumn{1}{c}{Term} & \multicolumn{2}{c}{Frequency}&
\multicolumn{1}{c}{Spectr.} & \multicolumn{1}{c}{S/N} &
\multicolumn{1}{c}{$\ell,m$} &
\multicolumn{1}{c}{Phot.}  \\
\multicolumn{1}{c}{}&  \multicolumn{1}{c}{[d$^{-1}$]} & \multicolumn{1}{c}{[$\mu$Hz]} &
\multicolumn{1}{c}{Ampl.} & \multicolumn{1}{c}{} &
\multicolumn{1}{c}{} & \multicolumn{1}{c}{Ampl.} \\
&&&&&& \multicolumn{1}{c}{[mmag]} \\
\noalign{\smallskip}
\hline
\noalign{\smallskip}
$f_{1}$     & 17.1616  &  198.628	     &0.245  &  14.8 &  1,0; 0,0   &  12.961 \\
$f_{11}$    & 15.0390  &  174.061	     &0.136  &   6.5 &  4,1   &   0.354\\
$f_{27}$    & 12.6917  &  146.894	     &0.136  &   9.7 &  6,-2  &   0.259\\
$f_{196}$   & 16.8485  &  195.005	     &0.133  &   7.3 &  6,5   &   0.059\\
$f_{203}$   & 16.4743  &  190.674	     &0.129  &   6.8 &  7,5   &   0.064\\
$f_{58}$    &  7.7266  &   89.428	     &0.113  &   8.5 &  7,-3  &   0.167\\
$f_{219}$   & 17.6884  &  204.726	     &0.128  &   7.9 &  6,4   &   0.051\\
$f_{28}$    & 11.7787  &  136.327	     &0.117  &   8.0 &  6,-5  &   0.257\\
$f_{79}$    &  6.5680  &   76.018	     &0.110  &   5.2 &  5,-2  &   0.125\\
$f_{8}$     & 15.1787  &  175.678	     &0.127  &   9.0 &  4,1   &   0.979\\
$f_{71}$    &  9.4454  &  109.321	     &0.101  &   5.9 &  7,-5  &   0.125\\
$f_{9}$     & 17.2805  &  200.005	     &0.076  &   5.9 &  3,1   &   0.903\\
$f_{115}$   & 15.3171  &  177.280	     &0.075  &   6.5 &  6,0   &   0.095\\
$f_{29}$    &  8.2442  &   95.418	     &0.060  &   5.2 &  9,-5  &   0.262\\
$f_{24}$    &  8.8886  &  102.877	     &0.057  &   3.0 &  8,-1  &   0.308\\
$f_{2}$     & 13.0499  &  151.040	     &0.073  &   5.0 &  2,0 (1,0)   &	2.342 \\  
$f_{83}$    & 10.6011  &  122.697	     &0.054  &   3.4 &  7,-3  &   0.112\\
$f_{218}$   & 16.9705  &  196.417	     &0.058  &   4.7 &  8,7   &   0.046\\
$f_{110}$   & 12.5337  &  145.065	     &0.052  &   3.6 &  7,1   &   0.094\\
\noalign{\smallskip}
\hline
\label{spectro}
\end{tabular}
\end{table}

\begin{figure}
\begin{center}
\includegraphics[width=\columnwidth,height=\columnwidth]{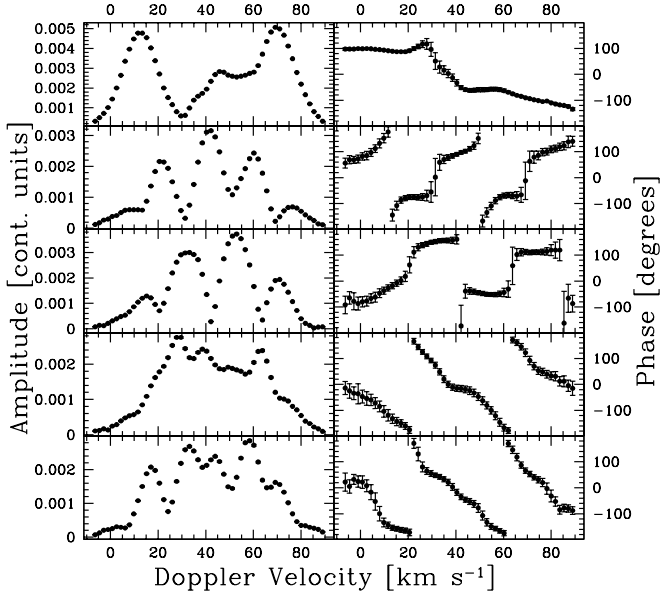}
\caption{\footnotesize
 From top to bottom: amplitude (left panel) and phase (right panel) variations across the line profile
of one axisymmetric
(17.1617~d$^{-1}$), two retrograde (7.7262 and 12.6917~d$^{-1}$), and two prograde modes (16.848 and 16.874~d$^{-1}$).
}
\label{ampfas}
\end{center}
\end{figure}

The amplitude and phase variations of all spectroscopic modes run across the whole line profile
(from $-10$ to  90 km~s$^{-1}$, some examples are shown in Fig.~\ref{ampfas}):
this gives us the certainty that 
the pulsational variability has to be ascribed to the fast rotator.

The spectroscopic modes are shown in the bottom panel of Fig.~\ref{freqord}.
The pseudo-Nyquist frequency (21.06~d$^{-1}$) must be taken into account when comparing 
them with the photometric counterparts (top panel).
The amplitudes (the third column of Table~\ref{spectro}) are integrated on the whole LSD 
profile, while their S/Ns (fourth column)  are computed by FAMIAS.
The shape of amplitude and phase curves of the dominant
mode (17.1617~d$^{-1}$, top panel of Fig.~\ref{ampfas}) looks quite entangled across the line profile,
but it seems to be an axisymmetric mode.
This probably indicates the presence of
another unresolved mode superimposed on the axisymmetric one.
Any attempts to fit this peak taking into account
other nearby peaks (considering adjacent photometric detections)
did not supply significant improvements because of the complexity of the spectral 
region around $f_1$ (see Fig.~\ref{freqord}).

\subsection{Line profile moments}

An independent search for frequencies was performed on the timeseries defined by the $0^{\rm th}$ 
(equivalent width, EW), $1^{\rm st}$ (radial velocity), and  $2^{\rm nd}$ (line width)
order moments. For these computations the narrow absorption profile (approximated by a Gaussian) was  
subtracted, because the amplitudes of the moment variations are dependent on the average profile shape. 
Because the moments are integrated over the whole
stellar disc, they are more sensitive to low--degree modes.
The results are shown in Table~\ref{moments}.
We see that the first 15 photometric terms were independently detected at least in one of the moment curves.
In particular, all were detected in the radial velocity variations. 

\begin{table*}
\begin{flushleft}
\caption{Modes detected by the line moment variations. All were associated to a photometric term.}
\begin{tabular}{rrrrrrrrrr}
\hline
\hline
\noalign{\smallskip}
\multicolumn{1}{c}{Term}&
\multicolumn{2}{c}{Frequency} &
\multicolumn{2}{c}{EW}&
\multicolumn{2}{c}{Rad. vel.} &
\multicolumn{2}{c}{Line width} &
\multicolumn{1}{c}{Phot. Amp.}\\
\multicolumn{1}{c}{}&
\multicolumn{1}{c}{[d$^{-1}$]}&
\multicolumn{1}{c}{[$\mu$Hz]}&
\multicolumn{1}{c}{[$10^{-2}$ km~s$^{-1}$]} &
\multicolumn{1}{c}{S/N} & 
\multicolumn{1}{c}{[km s$^{-1}$]} &
\multicolumn{1}{c}{S/N}&
\multicolumn{1}{c}{[km s$^{-1}$]$^2$} & 
\multicolumn{1}{c}{S/N} & 
\multicolumn{1}{c}{[mmag]} 
\\
\noalign{\smallskip}
\hline
\noalign{\smallskip}
 $f_{79}$ &  6.5680  &   76.018&      &      &	  &    & 4.4   & 5.8 & 0.125  \\
 $f_7 $   &  7.1320  &   82.546&      &      &  0.08 & 6.3&       &     & 0.974  \\
 $f_{26}$ &  7.3304  &   84.842&      &      &  0.06 & 4.4& 3.7   & 4.8 & 0.280  \\
 $f_{10}$ &  7.6604  &   88.661&      &      &  0.04&  3.0&       &     &  0.824 \\
 $f_5 $   &  8.1366  &   94.173& 0.66 & 4.5  &  0.12&  9.4&       &     &  1.544\\
 $f_{24}$ &  8.8886  &  102.877& 0.47 & 3.5  &  0.05&  3.4& 3.7   & 5.8 &  0.308 \\
 $f_2$    & 13.0499  &  151.040& 1.15 &11.4  &  0.24& 17.0&       &     &  2.342 \\
 $f_{30}$ & 13.3778  &  154.835&	   &	  &      &     &  3.2  & 3.8 & 0.260   \\
 $f_6$    & 13.6605  &  158.107& 0.62 & 4.7  &  0.11&  7.9&       &     &  1.313 \\
 $f_{15}$ & 13.9725  &  161.718&	   &	  &  0.05&  3.9&       &     & 0.792   \\
 $f_{17}$ & 14.0608  &  162.740&	   &	  &      &     &   5.2 & 6.2 &  0.471  \\
 $f_{11}$ & 15.0390  &  174.061&	   &	  &  0.04&  3.0&   9.2 &10.8 &  0.836  \\
 $f_8 $   & 15.1787  &  175.678& 6.46 & 6.1  &  0.18& 13.7&  12.1 &14.0 &  0.979 \\
 $f_{14}$ & 15.8217  &  183.120& 0.45 & 4.3  &  0.07&  5.3&       &     & 0.707   \\
 $f_{13}$ & 15.8344  &  183.267&	   &	  &  0.14&  9.7&       &     &  0.726  \\
 $f_4$    & 16.0024  &  185.212& 0.55 & 5.2  &  0.20& 13.5&       &     &  1.593 \\
 $f_3$    & 16.2507  &  188.086& 1.04 & 9.6  &  0.16&10.9 &       &     & 1.736  \\
 $f_1 $   & 17.1617  &  198.630& 5.01 &43.9  &  1.23& 69.1&   5.2 & 6.9 & 12.961 \\
 $f_{12}$ & 17.1863  &  198.914&	   &	  &  0.12&  6.7&       &     &  0.692  \\
 $f_9$    & 17.2805  &  200.005&	   &	  &  0.11&  6.3&   6.0 & 8.0 &  0.903 \\
    \noalign{\smallskip}
\hline
\label{moments}
\end{tabular}
\end{flushleft}
\end{table*}

All modes detected in the zero-order moments were also  detected in the
first-order moment and in general correspond to the strongest photometric terms.
The only exception is the 8.8886~d$^{-1}$ mode, which probably  suffers from the
contamination of the 7.8843~d$^{-1}$ mode.
That several of the detected modes in the first-order moment were not
detected in the second-order one, and in particular among those with the strongest
photometric amplitudes, suggests that these could be axisymmetric modes ($m=0$).
The detection of axisymmetric modes is favoured for geometrical reasons because the star
is seen almost pole-on (see paragraph 3.5). 
We point out  $f_2$=13.0499, $f_3$=16.2507, $f_4$=16.0024, $f_5$=8.1366, and $f_6$=13.6605~d$^{-1}$ as
the most promising candidates.

\begin{figure}[]
\begin{center}
\includegraphics[width=\columnwidth,height=\columnwidth]{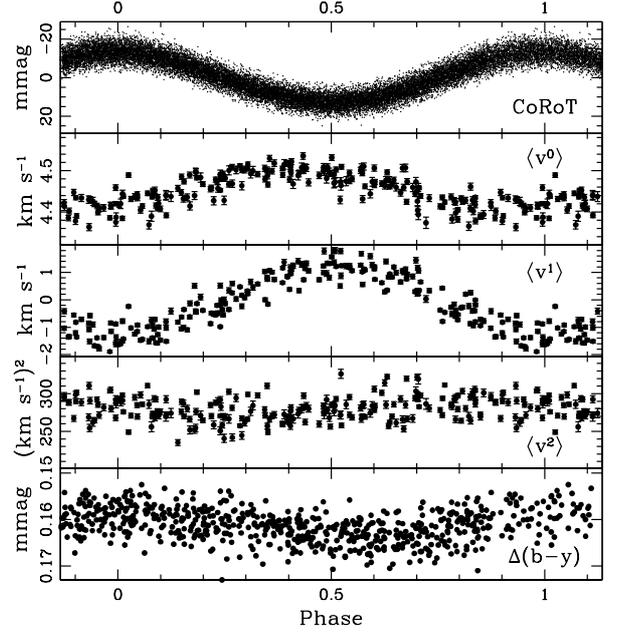}
\caption{\footnotesize From top to bottom: light, first three moments 
and $\Delta$(b-y) (comparison minus variable) 
variations folded in phase according to the main pulsation frequency $f_1$=17.1617~d$^{-1}$.}
\label{lm}
\end{center}
\end{figure}

The dominant photometric mode (17.1617~d$^{-1}$) was detected in all moment curves.
The variation of the first three moments phased according to the dominant period 
are compared with the light variations in Fig.~\ref{lm}.
However, while $f_1$  is the strongest term in the zeroth- and first-order moment, 
it is not the dominant term in the second-order one, where the dominant mode is 
$f_8$=15.1787~d$^{-1}$.
We cannot exclude that there may be a reciprocal partial contamination caused by aliasing,
considering that the distance between the two peaks is 1.983~d$^{-1}$.
Maybe this contamination could also explain the perturbed shapes of amplitude and phase variations across 
the line profile (Fig.~\ref{ampfas}, top panels).

\subsection{Mode identification}
It is difficult to accurately fix the stellar physical parameters, considering
all the uncertainties connected to the disentanglement of the spectra and the 
fact that the stellar parallax is unknown. However, if we consider 
the parameters given in Table~\ref{physpar} to be sufficiently reliable,
we can derive from the tables by \citet{strkur}
that the primary star (i.e. the $\delta$ Sct variable) has an A8 III spectral classification and therefore a
bolometric magnitude of 1.1$\pm$0.3~mag. Such a star would lie 
 practically in the centre of the $\delta$ Sct instability strip \citep{brevie}.
With Eq.~6 in \citet{brevie}  and using $\log g$ and $T_{\rm eff}$  from Table~\ref{physpar} 
and the above quoted M$_{\rm bol}$, we can estimate that the frequency of the  radial
fundamental mode (Q=0.033~d) should be about 7.4$\pm$2.2~d$^{-1}$. 

From the paper by \citet{schaller} we find, using the above derived  $T_{\rm eff}$ and M$_{\rm bol}$ values, 
 a mass of 2.1$\pm$0.3\,$M_{\odot}$.
Combining this with the estimated $\log g\sim 3.7$, we derive a radius of $3.4\pm1.2~R_{\odot}$,
and we adopted this value for the subsequent mode identification.
 An attempt to derive a more accurate value by applying
the Baade--Wesselink technique using light, colour ($b-y$ variations 
 are shown at the bottom of Fig.~\ref{lm}; see Sect.~3.6 for details) and radial velocity
variations of the dominant mode (17.17~d$^{-1}$) did not lead to a significant result.
In turn, this supports the $\ell=1$ identification for the $f_1$ mode.
Only for this  $\ell$ value the projected area 
variation is null and the Baade-Wesselink technique breaks down \citep{stamford}.

\begin{figure}[]
\begin{center}
\includegraphics[width=\columnwidth,height=\columnwidth]{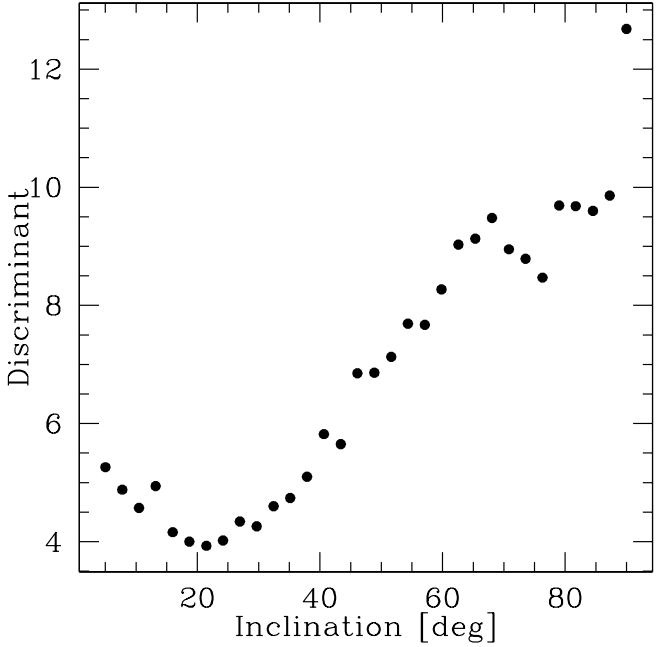}
\caption{\footnotesize 
Behaviour of the discriminant obtained by fitting  all 19
spectroscopic modes simultaneously to the LSD profiles. 
}
\label{incl}
\end{center}
\end{figure}

The mode identification was performed with FAMIAS in the AP mode \citep[see ][]{famias}. 
Since the profiles shapes are complex because of binarity, their average value was subtracted from the 
individual LSD profiles and the identification was performed on these residuals.
A first identification was performed individually on each detected mode by fitting 
its amplitude and phase variations 
across the line profile. This supplied preliminary $\ell$,$m$,$i$ values.
After that we performed a simultaneous fit of all modes with the previously detected $\ell$,$m$ values kept fixed, 
but leaving the inclination as free parameter.  
From the values of the discriminant  we were able to estimate  $i=21^{\rm o}\pm6^{\rm o}$
as the most probable inclination of the rotational axis (Fig.~\ref{incl}).
Finally, a separate fit was performed on each detected mode assuming this inclination value to improve 
the $\ell$,$m$ values.
The derived $\ell$,$m$ values are listed in Table~\ref{spectro}, where 
the negative $m$ values indicate retrograde modes.
Uncertainties are estimated to be $\pm$1 for the $\ell$ degree, while they are larger for the $m$ values.
The main cause for uncertainties are the contaminations by adjacent unresolved modes and by aliases.  
But the distinction between prograde and retrograde modes through the shape of the
 phase curves is quite unambiguous (see Fig.~\ref{ampfas}) and the total amplitude of these curves across
 the line profile supplies a good constraint to the $\ell$ value \citep{telting}.
 
Knowing the inclination, we were able to estimate v$_{eq}=116\pm38$~km~s$^{-1}$ from the $v\sin i$ value and
consequently the rotational frequency, $\nu_{rot}=0.67\pm0.32$~d$^{-1}$.
We applied the same approach to the secondary component and estimated that it could be an F2 III star, with
a bolometric absolute magnitude of 1.8$\pm$0.3~mag, 
located just outside the red border of the instability strip.

\subsection{Multicolour photometry}

We also observed HD 50870 in  Str\"omgren $uvby$ photometry
to have an independent tool to propose at least an
identification of the modes with the highest amplitudes.
$uvby$ photometry was performed at San Pedro
M\'artir Observatory from 2007 to 2010 and the
final dataset consists  of 593 datapoints collected on 28
nights for a total survey of 156.5~h. 
The error on a single measurement is 0.005~mag in the
$vby$ filters and around 0.015~mag in $u$ filter.
The frequency analysis of this dataset allowed us to
detect the terms from $f_1$ to $f_5$ in an unambiguous way. 
The aliasing effect is noticeable after the sixth term, since
instead of $f_6$=13.66~d$^{-1}$
the highest peak is  the alias at $f_6$+1~d$^{-1}$. The subsequent
residual spectrum became very noisy and the highest
peaks cannot be identified in  a reliable way
 with the frequencies detected in the CoRoT timeseries. 
Looking at the photometric amplitudes  listed in
Table~3, we infer that we were able to detect from ground all terms
with a CoRoT amplitude higher than 1.0~mmag. 

We calculated the amplitude ratios and phase shifts 
by using the $y$ colour as reference system \citep{garrido,por}. 
The predominant term $f_1$ is characterized by  
negative phase shifts (e.g., $\phi_v-\phi_y=-0.05\pm0.03$~rad,
$\phi_b-\phi_y=-0.04\pm0.03$~rad). The 
amplitude ratios are $A_v/A_y=1.39\pm0.04$ and $A_b/A_y=1.19\pm0.05$.
 Therefore, the case of HD~50870 is different from that of HD~50844,
where the positive phase shifts supported the identification of $f_1$
as radial mode  \citep[see Fig.~10 in ][]{por}. Here, the negative phase shifts
suggest a nonradial mode, but this identification is ambiguous 
since the $f_1$ mode might belong to a high radial order (we obtain Q=0.014~d
from the procedure described in Sect.~3.5) and
the theoretical phase shifts are negative in this case, too
\citep[see Fig.~13 in ][]{garrido}.
The phase shifts of the  terms from $f_2$ to $f_6$ 
also have negative values, but the error bars are at least three times larger than 
that  the $f_1$ ones, which precludes a reliable mode identification.

\section{Frequency spacings}  \label{multiplet}
The asymptotic regime of the solar-like oscillations describes 
the well-known periodic pattern characterized by the large 
\begin{equation}
\Delta\nu =\nu_{n+1,\ell}-\nu_{n,l}
\end{equation}
and the small 
\begin{equation}
\delta\nu=\nu_{n,\ell}-\nu_{n-1,\ell+2}.
\end{equation}
separations. This regime is not valid for $\delta$ Sct stars
and we were unable to detect the two separations as sharp structures in the frequency 
distributions; instead they appeared as quasi-regular spacings \citep{garcia}. 
These spacings become more evident when the search is restricted to the 
modes covering a small $\ell$ range. Because the modes with $\ell<5$ should show
the highest amplitudes, we performed our analysis by
considering the frequencies with amplitudes higher than 0.1~mmag.

\begin{figure}[]
\begin{center}
\includegraphics[width=\columnwidth,height=\columnwidth]{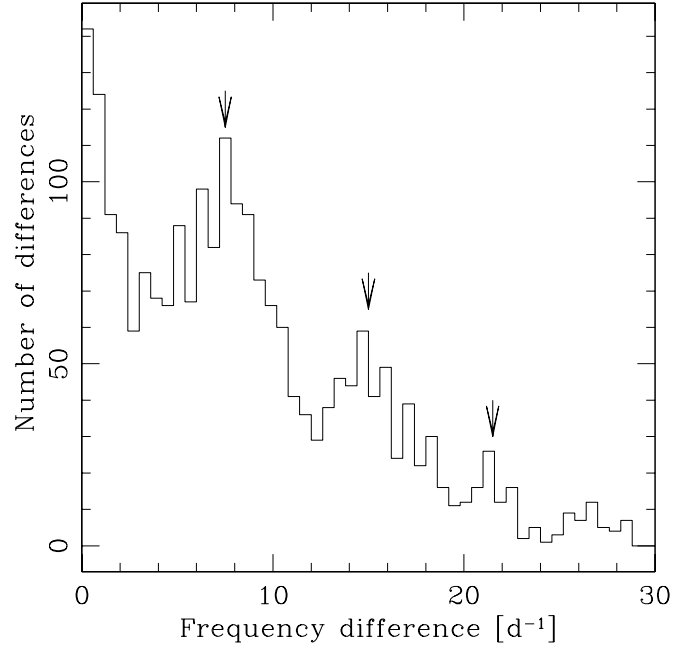}
\caption{\footnotesize Histogram of the differences between couples of 
frequencies with amplitudes
higher than 0.1~mmag in the 0-30~d$^{-1}$ interval. The pattern at 7.5, 15,
and 22~d$^{-1}$ is indicated.} 
\label{histogram}
\end{center}
\end{figure}

\begin{figure}[]
\begin{center}
\includegraphics[width=\columnwidth,height=\columnwidth]{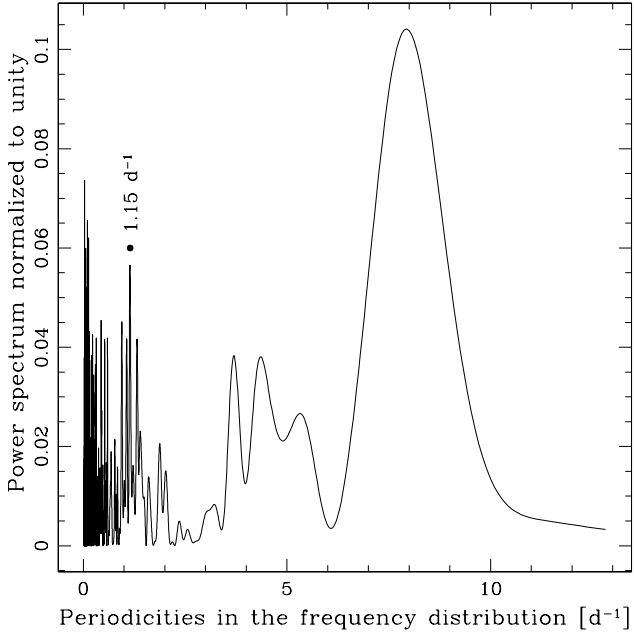}
\caption{\footnotesize Power spectrum derived from the distribution of the frequencies
with amplitudes higher than 0.1~mmag. 
The strongest peak is close to the first peak of the pattern shown in Fig.~\ref{histogram}.
The peak at 1.15~d$^{-1}$, close to twice the rotational frequency, is also
indicated. }
\label{ft50870}
\end{center}
\end{figure}

Searching for  the large  spacing (Fig.~\ref{histogram})
we found a prominent peak around 7.5~d$^{-1}$ and  
other peaks at 15~d$^{-1}$ and 22~d$^{-1}$,  i.e., about double and 
triple frequency of the former value.  This suggests a regular pattern with
a frequency separation of 7.5~d$^{-1}$.
 We also computed the Fourier power spectrum of a series of
Dirac's $\delta$ functions with the same amplitude and
centred on the frequencies detected in the CoRoT data.
This power spectrum is characterized by a broad peak centred at about 7.8~d$^{-1}$
(Fig.~\ref{ft50870}).

\begin{figure}[]
\begin{center}
\includegraphics[width=\columnwidth,height=\columnwidth]{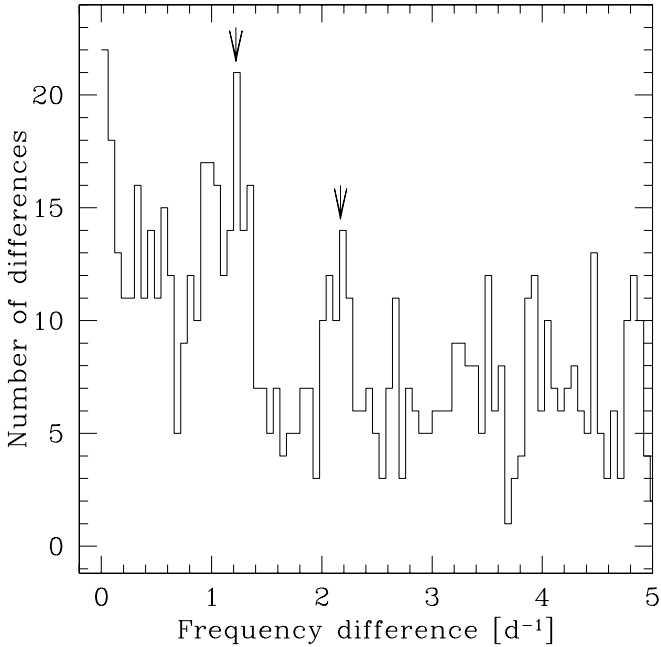}
\caption{\footnotesize Histogram of the differences between couples of
frequencies with amplitudes higher than 0.1~mmag in the 0-5~d$^{-1}$ interval.
The peak at 1.2~d$^{-1}$ and its double are visible.}
\label{histosplit}
\end{center}
\end{figure}

When searching for a smaller regular spacing  in the 0--5~d$^{-1}$ interval,
we were able to identify a prominent peak at about 1.2~d$^{-1}$ (Fig.~\ref{histosplit}). 
The 1.2~d$^{-1}$ value is approximately twice the rotational frequency 
(0.67~d$^{-1}$) and corresponds to the peak shown in Fig.~\ref{ft50870}.
The recurrency of the 0.6~d$^{-1}$ separation is confirmed  by some peaks
reproducing a multiplet pattern, as those observed at 6.70, 7.13, 7.66, 8.14, and 8.76 ~d$^{-1}$.

The study of the frequency spacings provided two main results.
First, the Fourier transform and the histogram in the 0-5~d$^{-1}$ range
(Fig.~\ref{ft50870} and Fig.~\ref{histosplit})
show the double of the rotational splitting, which is predicted 
for a frequency range high enough to render the effect of
the Coriolis force negligible \citep{lgn10}. 
Second, the frequency differences in the  5-30~d$^{-1}$ range
show a spacing periodicity  
\citep[as defined by ][]{garcia} of about 7.5~d$^{-1}$.

\section {Asteroseismic modelling}

The availability of the physical parameters and of the full description of the
pulsational spectrum allowed us to  perform
the computation of asteroseismic models (i.e., equilibrium plus pulsation models), the analysis
of instability,  and a study of quasi-periodicities in the observed oscillation spectrum
\citep{garcia}.

\subsection{The asteroseismic models} \label{sec51}

Equilibrium models were computed with the evolutionary code {\sc cesam} \citep{Morel97, MorelLeb08}. 
Physics included in the models was appropriate to the description of  $\delta$ Sct stars \citep{casas}.
We considered a variation in the convective efficiency parameter $\alpha_{MLT}$ up to 1.5, of the overshoot up to 0.3,
and  a solar metallicity. 
Diffusion and radiative forces were assumed to be negligible for this type of 
stars \citep{Goupil05} and were not included in the modelling. We considered about 2000 points for
the computation mesh, following the studies of the Evolution and Asteroseismic Tools Activities 
\citep[ESTA/CoRoT\footnote{http://www.astro.up.pt/corot/}; ][]{Moya08esta, Lebreton08esta}. 

We considered $i=21^{\rm o}\pm6^{\rm o}$ and then $v_{\rm eq}=116\pm38\,$km~s$^{-1}$, 
as resulting from the spectroscopic mode identification.
For these rotational velocities it is expected to find a non-rigid rotation in the
stellar interior, in particular at the interface between the convective core and the radiative envelope. 
In that region the mixed modes (low-order $p$ and $g$ modes) have a significant sensitivity 
to variations of the internal rotation profile \citep{nueri}. 
Without a detailed description of the shape of the internal rotation profile, 
we can reasonably approximate it with 
the physical assumption of a local conservation of the angular momentum during the evolution 
\citep{Sua06rotcel} in the radiative zones, and a
quasi-instantaneous transport of the angular momentum in the convective zones. The total angular momentum
is conserved (no mass loss is expected during the main-sequence stage). 

Pulsational models were obtained by a perturbation of the equilibrium 
models using the adiabatic oscillation code {\sc filou} \citep[][ and references therein]{phd}. 
Models were computed using the physical parameters obtained from HARPS spectroscopy
(see Table~\ref{physpar}). The uncertainties in those parameters defined our space of validity 
for the representative models. 
Models were constrained to predict the highest amplitude modes having the expected spherical 
degrees (Table~\ref{spectro}).
This procedure defined two sets of models with masses of $M$=2.10 and 2.18\,$M_{\odot}$.  
\begin{figure}[]
\begin{center}
\includegraphics[width=\columnwidth,height=\columnwidth]{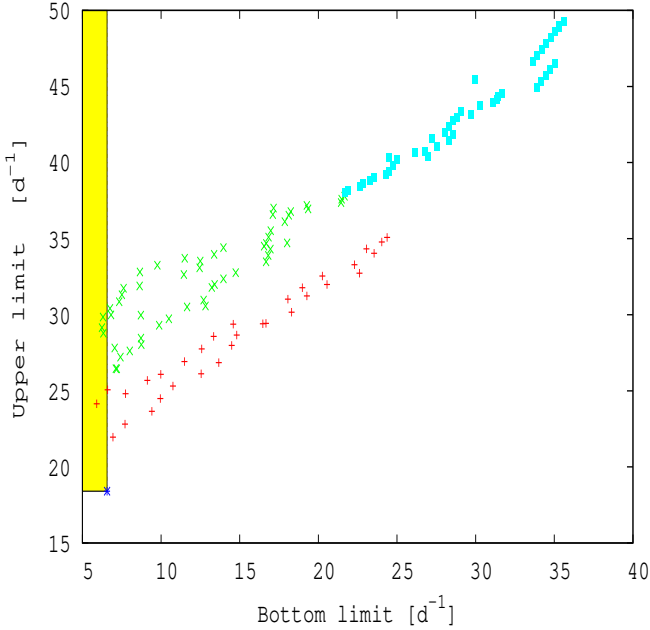}
\caption{\footnotesize 
Bottom versus upper limit of the instability range.  
Models computed using the physical parameters from spectroscopy are plotted as red and green crosses.
The limits of the frequencies with the highest amplitudes observed in HD\,50870 are depicted with a blue star.
The shaded yellow area shows where a theoretical model must be to predict an
 over-stable frequency range containing the observed one.
The models with a $\Delta\nu$ in the observed range of [3.5,4.5]~d$^{-1}$ are marked as light blue
squares.}

\label{insta}
\end{center}
\end{figure}

\subsection{The instability analysis}

A non-adiabatic analysis of the mode energy balance of the sets of models was performed with
 the GraCo code \citep{moya1,moya2}. 
Theoretical results were compared with the range covered by the spectroscopic
modes and by most of the highest amplitude photometric modes, i.e., [6.5,18.4]~~d$^{-1}$
(Tables~\ref{comb} and  \ref{spectro}).
Each  theoretical model is represented by a point in Fig.~\ref{insta}. 
We found a  few models that predicted the over-stable region [6.5,18.4]~d$^{-1}$. 
The analysis of instability thus constrains the representative models to  a first set of models with 
$M$=2.18\,$M_{\odot}$,  3.8692$<\log\,T_{\rm eff}<$3.8811, 3.7003$<\log\,g<$3.7746, 1.5009$<L/L_{\odot}<$1.5056 
and a second set with 
$M$=2.10\,$M_{\odot}$,  3.8673$<\log\,T_{\rm eff}<$3.8803, 3.7935$<\log\,g<$3.8529, 1.3812$<L/L_{\odot}<$1.3886. 

\subsection{Analysis of quasi-periodic patterns}

Using the aforementioned sets of theoretical models, we looked for 
quasi-periodic patterns. There is no model with a mean $\Delta\nu$ 
around 7.8~d$^{-1}$ (Fig.~\ref{ft50870}).
The pulsational models taking into account the rotational effects
up to the second order of the perturbative approach did not point out
structures similar to the multiplet centred at 7.66~d$^{-1}$ (Sect.~\ref{multiplet}).
Consequently, we considered the half value of 3.9~d$^{-1}$.  
We estimated the uncertainty of this value using the FWHM of the highest peak of the Fourier 
transform (Fig.~\ref{ft50870}). Thus, we obtained a [3.5,4.5]~d$^{-1}$ range for $\Delta\nu$. 
Models with 2.10~$M_{\odot}$ show these values, but
they also  predict a  lower limit for the excited modes that is much higher than the observed one
(Fig.~\ref{insta}). 

Therefore, there is no model that at the same time predicts the [6.5,18.4]~d$^{-1}$ range of observed frequencies
and the [3.5,4.5]~d$^{-1}$ range of the quasi-periodicity intepreted as a large separation. 
This seems a general problem in the era of rich pulsational spectra detected
by space observations, as stressed by  \citet{katrien11} when analysing 
the $\delta$ Sct stars observed with {\it Kepler}. 

\section{Discussion and conclusions}

The space-based photometric  and the ground-based spectroscopic monitorings
of another $\delta$ Sct star after HD 50844 \citep{por} confirm
that these variables show a complicated amplitude spectrum. 
HD~50870  has  one predominant pulsation mode
(17.1617 d$^{-1}$, 12.96~mmag), and about 20 modes with amplitudes beween 2.3 and 0.3~mmag and
frequencies between 7 and 19~d$^{-1}$. Well-isolated peaks were detected down to 0.05~mmag.
The frequencies  with amplitudes lower than about 50~$\mu$mag (S/N$\sim$12) form a flat low--frequency
plateau with a cut--off at about 35d$^{-1}$ ($\sim$400~$\mu$Hz).
After the detection of  several hundreds of frequencies (down to  
amplitudes of about 12~ppm),  
there is still an excess of power at low frequencies
(see last panel of Fig. 4). About 1800  frequency terms are necessary to obtain a flat power spectrum
by means of  a classical Fourier solution of the light curve. 

\begin{figure}[]
\begin{center}
\includegraphics[width=\columnwidth,height=\columnwidth]{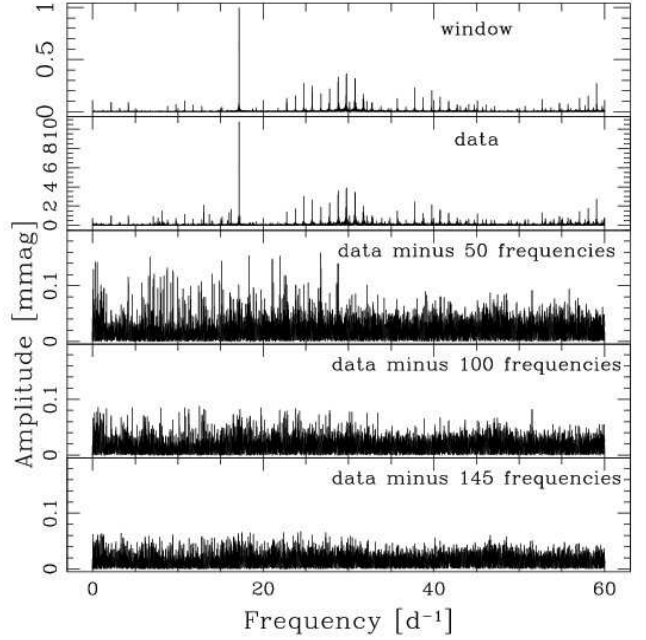}
\caption{\footnotesize 
Frequency analysis of the HD~50870 light curve but sampled with the data binning of the Kepler satellite.
Only 145 peaks are detectable above the S/N level of 4.0. The spectral window shifted at the frequency of the
predominant mode (17.16~d$^{-1}$) is shown in the top panel.}
\label{kepler}
\end{center}
\end{figure}

The results obtained for HD~50870 confirm that we need hundreds of frequencies to reproduce
the light curves of $\delta$ Sct stars, i.e., that HD~50844 is not an exception or a unique 
case \citep{baldzi}. Because the light curves of $\delta$ Sct stars show rapid variations, 
the detailed investigation of the signal content must be performed on the basis of time series 
with a short cadence sampling. To illustrate the point, we grouped the CoRoT measurements of HD~50870
in bins of 30~min, i.e., the typical long cadence of the {\it Kepler} data. The resulting amplitude
spectrum obtained from the frequency analysis of the 5135 datapoints is shown in Fig.~\ref{kepler}. 
The spectrum is generally flat and a limited set of frequencies (145) is above the significance
limit of 4.0. This can be explained by the fact that the pseudo--Nyquist frequency of the almost equally--spaced
{\it Kepler}--like data is about 23~d$^{-1}$. Consequently, a consistent fraction of the spectral power below
this limit (where most of the stellar signal is concentrated) is reflected above it, 
thus simulating a higher noise level. Indeed, we calculated a white noise level
of 15~$\mu$mag (to be compared to the 3~$\mu$mag level obtained from the analysis with the CoRoT
 data sampling).
We can conclude that the cadence of 30~min is unable to describe the
light variation of $\delta$ Sct stars in a complete way.

\begin{figure}[]
\begin{center}
\includegraphics[width=\columnwidth,height=\columnwidth]{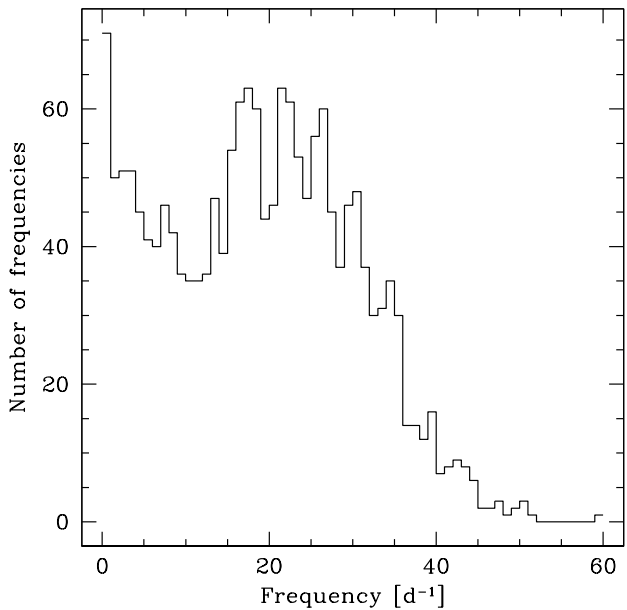}
\caption{\footnotesize 
Density distribution of the 1800 frequencies necessary to whiten the power spectrum 
(number of frequencies per d$^{-1}$) as a function of frequency.}
\label{histo}
\end{center}
\end{figure}

We remain with the still unsolved problem whether the excess of power detected in the
CoRoT timeseries below 50~d$^{-1}$ is due 
to the excitation of a huge number of modes or is the result of a 
broad--band low--frequency noise process, i.e.,   we are observing a discrete or continuum power spectrum. 
A similar plateau has been observed by \citet{por} in HD 50844 and it seems to be present
in HD 174936, too \citep{garcia}. The analysis  of the light variations of
other $\delta$ Sct stars observed by CoRoT seems to support this result: HD~170699, for instance,
which is a very fast rotator ($v\sin i$=270~km~s$^{-1}$) has a plateau with the cut--off
at about 25~d$^{-1}$ (Mantegazza et al., in preparation). 
A tentative preliminary answer can be given for HD 50870. 
 The histogram that shows the distribution with a binning 
of 1~d$^{-1}$ of the frequency terms needed to whiten the power 
 spectrum (i.e., to lower its mean value at the low frequencies at about the same level as above 60~d$^{-1}$)
 is shown in Fig.~\ref{histo}. Excluding the peak close to zero frequency, which might be affected by the
residual satellite low-frequency noise, the maximum density is 63 frequencies per ~d$^{-1}$, with an
average value below $\sim$ 35~d$^{-1}$ of about 50~frequencies  per ~d$^{-1}$. 
It is well known that if we have a time series generated by a process with a continuum power spectrum,
and we estimate its power spectrum by means of the discrete Fourier transform, 
we approximate it with a discrete spectrum
consisting of a bunch of peaks with a separation between them of the order of the reciprocal of the time series
baseline \citep[e.g.,][]{chatfield}.
The baseline of the CoRoT observations is  about 114~d, 
then we should expect a frequency density of the order of 114 frequencies~per~d$^{-1}$ to be detected
if there is a signal with a continuum power spectrum. 
 Since the observed density is decidedly lower,
this analysis rules out the possibility that the plateau is due to a 
process with a continuum power spectrum. However, it does not rule out   
the action of the granulation effect, since 
this effect could be described as a random process with
characteristic time scales that  could generate different densities.
The observed frequency ranges, amplitudes, and densities
pose observational constraints to the models of the convective layers in $\delta$ Sct stars, at the
moment limited to temperatures cooler than that of HD~50870 \cite[$T_{\rm eff}<$6850\,K, ][]{reza1}. 
A more detailed theoretical investigation
is necessary to support the effectiveness of convection in $\delta$ Sct stars.

As for HD~50844, high--resolution spectroscopy was of paramount importance to improve
the physical scenario. 
The analysis of the line profile variations allowed us to identify
nineteen progrades, axisymmetric and retrogrades modes having  $\ell\le$9 as well as 
an inclination angle of about 21$^{\rm o}$. The corresponding rotational velocity is about 
116~km~s$^{-1}$, and therefore HD~50870 is rotating at about 34\% of the break-up velocity. 

HARPS spectroscopy suggests that the predominant mode is axisymmetric ($m$=0), 
but it does not allow us to distinguish between a radial or a nonradial mode.
The greater difference between  a radial mode ($\ell=0$) and an axisymmetric nonradial one 
with $\ell=1$ is in the equatorial region, but in the current case of a star seen almost pole--on
this region supplies a very limited contribution  through the Doppler effect to the
overall line profile variations.
On the other hand, the Baade--Wesselink technique (Sect.~3.5) and
the amplitude ratios and phase shifts in different photometric
colours (Sect.~3.6) seem to indicate a nonradial mode.
More in general, despite the huge observational effort,
there is still a mismatch between the observational contraints and the 
theoretical modelling. The mechanism of selecting the modes with the
highest amplitudes is  not yet fully understood, also when considering
the effect of fast or moderate rotation.

\citet{antoci} observed a 0.1-0.2~mmag level of the solar--like oscillations
for HD~187547. 
We did not find any evidence of the presence of oscillations like this in 
HD~50870,
at least above the level of 12~ppm. The non-detection may be the result
of the dilution of the structures broadened by the mode lifetime into the
noise and an apparent reduced amplitude due to the light from the companion.
In both cases, the damping should be very effective to produce a null
detection. HD~187547 has 
physical parameters very similar to those of HD~50870,
but also a slight overabundance of some chemical elements.
Since the amplitude of solar-like oscillations depends on the surface metal abundance
\citep{reza2,reza3}, this particularity can explain why the  visibility of these
oscillations is enhanced in HD~187547. 
A well-hidden cooler companion could also be an explanation for the case of HD~187547
because we were able to detect the binarity of HD~50870 only thanks to the continuous
monitoring with high-S/N HARPS spectra.

\begin{acknowledgements}
The HARPS data are being obtained as part of the ESO Large Programmes
LP~182.D-0356 and LP~185.D-0056 (PI.: E.~Poretti). 
Mode identification results were
obtained with the software package FAMIAS developed in the framework of the 
FP6 Coordination Action Helio- and 
Asteroseismology (HELAS; http://www.helas-eu.org/). 
EP  thanks J.~Ballot, F.~Ligni\`eres, and M.~Pasek for enlightning discussions
during his stay at IRAP, Toulouse.
We thank J.~Vialle for checking the English form of the original manuscript.
PJA acknowledges financial support from grants AYA2009 -08481-E and
AYA2010-14840 of the Spanish Ministry of Science and Innovation (MICINN).
AM acknowledges the funding of AstroMadrid (CAM S2009/ESP-1496) and the
Spanish grants ESP2007-65475-C02-02, AYA 2010-21161-C02-02. 
JCS acknowledges the financial support from the Spanish Ministry of Science through its
{\it Plan Nacional del Espacio} under project AYA2010-12030-E and AYA2010-20982-C02-01.
This work was supported by the Italian PRIN-INAF 2010
{\it Asterosesismology: looking inside the stars with space- and
ground-based observations}.
\end{acknowledgements}


\begin{thebibliography}{}

\bibitem[Antoci et al.(2011)]{antoci} Antoci, V., Handler, G., Campante, T.L., et al. 2011, Nature, 477, 570 
\bibitem[Auvergne et al.(2009)]{flight} Auvergne, M., Bodin, P., Boisnard, L., et al. 2009, \aap, 506, 411
\bibitem[Baglin et al.(2006)]{baglin} Baglin, A., Auvergne, M., Barge, P., et al. 2006, ESA-SP, 1306,33
\bibitem[Balona \& Dziembowski(2011)]{baldzi} Balona, L.A., \& Dziembowski, W.A. 2011, \mnras, 417, 591
\bibitem[Breger(2000)]{brevie} Breger, M. 2000, 
in ``Delta Scuti and Related Stars", M.~Breger \& M.H.~Montgomery Eds., ASP Conf. Series, 210, 3
\bibitem[Breger et al.(2005)]{fgvir} Breger, M., Lenz, P., Antoci, V., et al.  2005, \aap, 435, 955
\bibitem[Casas et al.(2006)]{casas} Casas, R., Su\'arez, J.C., Moya, A., \& Garrido, R. 2006, \aap, 455, 1019
\bibitem[Chatfield(1984)]{chatfield} Chatfield, C. 1984, ``The analysis of Time Series: An Introduction", Chapman and Hall, 
London-New York
\bibitem[Daszynska-Daszkiewicz et al.(2006)]{dasdas} Daszynska-Daszkiewicz, J., Dziembowski, W.A., \& Pamyatnykh, A.A. 2006,  MemSAIt, 77, 113
\bibitem[Deeming(1975)]{deeming} Deeming, T.J. 1975, \apss, 36, 137
\bibitem[Donati et al.(1997)]{lsd} Donati, J.-F., Semel, M., Carter, B.D., Rees, D.E., \& Collier Cameron,
A. 1997, \mnras, 291, 658
\bibitem[Garrido(2000)]{garrido} Garrido, R.   2000, in ``Delta Scuti and Related Stars",
ed. M.~Breger \&  M.H.~Montgomery, ASP Conf. Series, 210, 67
\bibitem[Garc\'ia Hern\'andez et al.(2009)]{garcia}  Garc\'ia Hern\'andez, A., Moya, A., Michel, E., et al. 2009, \aap, 506, 79
\bibitem[Goupil et al.(2005)]{Goupil05} Goupil, M.-J., Dupret, M.A., Samadi, R. et al. 2005, J. Astrophys. Astr.,
26, 249
\bibitem[Ilijic(2004)]{cres}Ilijic, S. 2004,  ASP Conf. Series, 318, 107
\bibitem[Ilijic et al.(2004)]{ihpf}Ilijic, S., Hensberge, H., Pavloski, K. \& Freyhammer, L.M. 2004, ASP Conf. Series 318, 111
\bibitem[Kallinger \& Matthews(2010)]{kama} Kallinger, T., Matthews, J.M. 2010, \apj, 711, L35
\bibitem[Kupka et al.(2000)]{kupka} Kupka, F.G., Ryabchikova, T.A., Piskunov, N.E., et al. 2000, 
Baltic Astron., 9, 590
\bibitem[Lebreton et al.(2008)]{Lebreton08esta} Lebreton, Y., Monteiro, M.J.P.F.G., Montalb{\'a}n, J. et al.
2008, \apss, 316, 1
\bibitem[Lenz \& Breger(2005)]{P04} Lenz, P., \& Breger, M. 2005, CoAst, 146, 53
\bibitem[Ligni\`eres \&  Georgeot(2009)]{lgn09} Ligni\`eres, F., \&  Georgeot, B. 2009, \aap, 500, 1173
\bibitem[Ligni\`eres  et al.(2010)]{lgn10} Ligni\`eres, F.,  Georgeot, B., \& Ballot, J. 2010, AN, 331, 1053
\bibitem[Mantegazza(2000)]{manvie} Mantegazza, L. 2000, 
in ``Delta Scuti and Related Stars", M.~Breger \& M.H.~Montgomery Eds., ASP Conf. Series, 210, 138
\bibitem[McCuskey(1956)]{mccus} McCuskey, S.W. 1956, \apjs, 2, 271
\bibitem[Moya et al.(2008)]{Moya08esta} Moya, A., Christensen-Dalsgaard, J., Charpinet, S., et al.
 2008, \apss, 316, 231
\bibitem[Moya \& Garrido(2008)]{moya2} Moya, A. \& Garrido, R. 2008, Ap\&SS, 316, 129
\bibitem[Moya et al.(2004)]{moya1} Moya, A., Garrido, R., \& Dupret, M.~A. 2004, \aap,  414, 1081
\bibitem[Morel(1997)]{Morel97} Morel, P. 1997, \aaps, 124, 597 
\bibitem[Morel \& Lebreton(2008)]{MorelLeb08} Morel, P., \& Lebreton, Y. 2008, \apss, 316, 61
\bibitem[Piskunov et al.(1995)]{vald} Piskunov, N.E., Kupka, F., Ryabchikova, T.A., et
al. 1995, \aaps, 112, 525
\bibitem[Poretti et al.(2005)]{gaudi} Poretti, E., Alonso, R., Amado, P.J., et al. 2005, \apj, 129, 2461
\bibitem[Poretti et al.(2009)]{por} Poretti, E., Michel, E., Garrido, R., et al.
2009, \aap, 506, 85
\bibitem[Rainer(2003)]{monica} Rainer, M. 2003, Laurea Thesis (in Italian), Universit\`a degli
Studi di Milano
\bibitem[Reegen(2007)]{sigspec} Reegen, P. 2007, \aap, 467, 1353
\bibitem[Samadi et al.(2002)]{reza1} Samadi, R., Goupil, M.-J., \& Houdek, G. 2002, \aap, 395, 563 
\bibitem[Samadi et al.(2010a)]{reza2} Samadi, R., Ludwig, H.-G., Belkacem, K., Goupil, M.-J.,  \& Dupret,
M.-A.  2010a, \aap, 509, A15 
\bibitem[Samadi et al.(2010b)]{reza3} Samadi, R., Ludwig, H.-G., Belkacem, K.,  et al. 2010b, \aap, 509, A16
\bibitem[Schaller et al.(1992)]{schaller} Schaller, G.,Schaerer, D., Meynet, G., \& Maeder, A. 1992, \aaps, 96, 269
\bibitem[Stamford \& Watson(1981)]{stamford} Stamford, P.A. \& Watson, R.D. 1981, \apss, 77, 131
\bibitem[Straizys \& Kuriliene(1981)]{strkur} Straiziys, V. \& Kuriliene, G. 1981, \apss, 80, 353
\bibitem[Stuetz et al.(2006)]{atc} Stuetz, Ch., Bagnulo, S., Jehin, E., et al. 2006, \aap, 451, 285
\bibitem[Su\'arez(2002)]{phd} Su{\'a}rez, J. C. 2002, Ph.D. Thesis, ISBN 84-689-3851-3, ID
02/PA07/7178
\bibitem[Su\'arez et al.(2006)]{Sua06rotcel} Su{\'a}rez, J.~C., Goupil, M.~J., \& Morel, P. 2006,
\aap, 449, 673
\bibitem[Su\'arez et al.(2009)]{nueri} Su{\'a}rez, J.~C., Moya, A., Amado, P.J., et al. 2009,
\apj, 690, 1401
\bibitem[Telting \& Schrijvers(1997)]{telting} Telting, J.H., \& Schrijvers, C. 1997, \aap, 317, 723
\bibitem[Uytterhoeven et al.(2011)]{katrien11} Uytterhoeven, K., Moya, A., Grigahc{\`e}ne, A., et al.
2011, \aap, 534, A125
\bibitem[Valenti \& Piskunov(1996)]{sme} Valenti, J.A., \& Piskunov, N. 1996, \aap, 118, 595
\bibitem[Zima(2008)]{famias} Zima, W. 2008, CoAst, 155, 17

\end{thebibliography}
\end{document}